\documentclass[12pt]{article}
\usepackage{comment}
\usepackage{amsmath}
\usepackage{amssymb}
\usepackage{graphicx}
\usepackage{here}
\usepackage{ascmac}

\usepackage{subcaption}
\usepackage{url}
\usepackage[sort&compress, numbers, merge]{natbib}
\usepackage{braket}
\usepackage[compat=1.1.0]{tikz-feynhand}
\usepackage{slashed}
\usepackage{mathtools}
\usepackage{todonotes}
\usepackage{framed}
\usepackage[colorlinks=true, linkcolor=blue, citecolor=blue,urlcolor=blue]{hyperref} 

\graphicspath{{./figures/}} 

\numberwithin{equation}{section}
\newcommand{\Slash}[1]{{\ooalign{\hfil/\hfil\crcr$#1$}}}

\newcommand{\Mpl}{M_{\rm Pl}}
\newcommand{\mysigma}{\sigma}

\newcommand{\red}[1]{{\color{black} #1}}

\begin{document}

\begin{titlepage}

\begin{center}
YITP-23-57,
KOBE-COSMO-23-05,
UT-Komaba/23-2,
IPMU23-0012, 
CTPU-PTC-23-13
\end{center}

\vskip 1.1cm

\begin{center}

{\Large \bf 
Gravitational Positivity for Phenomenologists: \\
\bigskip
Dark Gauge Boson in the Swampland
}

\vskip 1.2cm

Katsuki Aoki$^a$,
Toshifumi Noumi$^{b,c}$,
Ryo Saito$^{d,e}$,
Sota Sato$^{b,c}$, \\
Satoshi Shirai$^{e}$, 
Junsei Tokuda$^f$,
Masahito Yamazaki$^{e,g}$
\vskip 0.5cm

{\it

$^{a}${Center for Gravitational Physics and Quantum Information, \\
Yukawa Institute for Theoretical Physics, Kyoto University, 606-8502, Kyoto, Japan}

$^{b}$Graduate School of Arts and Sciences, University of Tokyo, \\
Komaba, Meguro-ku, Tokyo 153-8902, Japan

$^{c}${Department of Physics, Kobe University, Kobe 657-8501, Japan} 

$^{d}${Graduate School of Sciences and Technology for Innovation, \\ 
Yamaguchi University, Yamaguchi 753-8512, Japan}

$^{e}${Kavli Institute for the Physics and Mathematics of the Universe (WPI), \\
University of Tokyo Institutes for Advanced Study, \\ 
University of Tokyo, Kashiwa 277-8583, Japan}

$^{f}$Particle Theory and Cosmology Group, Center for Theoretical Physics of the Universe,\\ 
Institute for Basic Science (IBS), Daejeon, 34126, Korea

$^{g}${Trans-Scale Quantum Science Institute, \\
University of Tokyo, Tokyo 113-0033, Japan}
}

\vskip 1.0cm
\abstract{
The gravitational positivity bound gives quantitative ``swampland'' constraints on low-energy effective theories inside theories of quantum gravity. We give a comprehensive discussion of this bound for those interested in applications to phenomenological model building. We present a practical recipe for deriving the bound, and discuss subtleties relevant for realistic models. As an illustration, we study the positivity bound on the scattering of the massive gauge bosons in the Higgs/St\"{u}ckelberg mechanism. Under certain assumptions on gravitational amplitudes at high energy, we obtain a lower bound $m_{V} \gtrsim  \Lambda_\mathrm{UV}^2 /g M_\mathrm{Pl}$ on the gauge boson mass $m_V$, where $g$ is the coupling constant of the gauge field, $M_\mathrm{Pl}$ is the reduced Planck mass and $\Lambda_\mathrm{UV}$ is the ultraviolet cutoff of the effective field theory. This bound can strongly constrain new physics models involving a massive gauge boson. \red{We also discuss how the bound depends on our high-energy assumptions.}
}
\end{center}

\end{titlepage}

\tableofcontents

\section{Introduction} \label{sec:introduction}

It is one of the outstanding questions in present-day physics to uncover the origin and the identity of the dark matter (DM). More broadly we are interested in the search for the dark sector containing DM, or any physics beyond the Standard Model (BSM).

Traditionally, dark sectors have been analyzed in the framework of low-energy Effective Field Theories (EFTs), where the effects of the dark sector have been incorporated by small couplings of the dark sector to 
 particles of the Standard Model (SM). It is often the case that such couplings are
severely constrained by experiments/observations, leading to very small (absolute) values of parameters. While certain fine-tunings are needed,
the dominant attitude has been that such small parameters are well-tolerated in the EFTs, as long as such small parameters are technically natural~\cite{tHooft:1979rat}. This raises a fundamental problem for dark-sector searches: while we can keep eliminating parameter spaces for dark-sector-SM couplings, there is in principle no lower bound on their size, and it seems that we can never exclude a  ``nightmare'' scenario where 
the dark sector interacts with the SM sector only through gravitational interactions.

The situation is different, however, once we impose an extra condition that the Infrared (IR) low-energy EFT has a consistent ultraviolet (UV) completion with gravity. There has been mounting evidence recently that 
there are necessary conditions for a low-energy EFT to have a consistent UV completion, and 
such conditions are often called swampland constraints in the literature \cite{Vafa:2005ui, Ooguri:2006in}. 
One way to obtain such constraints is to invoke the so-called positivity constraints on scattering amplitudes at low energy. 
The positivity constraints are well established in non-gravitational theories and their application to the EFT of the Standard Model and cosmology has long been discussed: see \cite{Pham:1985cr,Pennington:1994kc,Ananthanarayan:1994hf,Comellas:1995hq} for early references on QCD. 
For more recent developments, see e.g.  \cite{Adams:2006sv}, a review article \cite{deRham:2022hpx} and references therein.
The advantage of this method is that we can derive an infinite set of inequalities on the EFT couplings by minimal assumptions on the scattering amplitudes such as unitarity, analyticity, and locality, without relying too much on the details of the UV completion. 

    \begin{figure}[t!]
        \centering
    \begin{screen}
    \begin{enumerate}
    \item [0.] Write down the (renormalizable) Lagrangian of your model, couple the model to gravity, and consider a $2\to 2$ scattering process $AB \to AB$.
        \item [1.] Compute the amplitude ${\cal M}_{\text{non-grav}}$ of diagrams  without gravity:
    \begin{align}
        i{\cal M}_{\text{non-grav}} (s,t)=
     \raisebox{-14pt}{\includegraphics{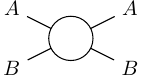} } + \cdots  \, , \notag 
    \end{align}
    and ${\cal M}_{\text{grav},t\text{-channel}}$  of graviton $t$-channel diagrams:
    \begin{align}
        i{\cal M}_{\text{grav},t\text{-channel}} (s,t) =
     \raisebox{-16pt}{\includegraphics{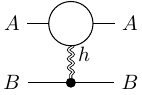}} + \cdots  \, . \notag
    \end{align}
      \item [2.] 
      Compute $B_{\text{non-grav}}$:
      \begin{align}
       B_{\text{non-grav}}(\Lambda) =\frac{4}{\pi}\int_{\Lambda^2}^{\infty} ds\, \frac{{\rm Im}\,\mathcal{M}_{\text{non-grav}}(s,t=0)}{\left(s-m_A^2 - m_B^2\right)^3}     \, . \notag
      \end{align}
       \item [3.] 
      Compute series expansion of ${\cal M}_{\text{grav},t\text{-channel}}$ around $t=0$ and $s = m_A^2 + m_B^2$, and extract $B_{\text{grav}}$ from the coefficient of $t^0 (s - m_A^2 -m_B^2)^2$.
        \item [4.] 
        Compute $B(\Lambda) = B_{\text{non-grav}}(\Lambda) +B_{\text{grav}} (\Lambda)$ and check  its positivity.
          \end{enumerate}
    \end{screen}
        \caption{Workflow to get the gravitational positivity bound (scalar case).}
        \label{fig:flow}
    \end{figure}

It is challenging to derive positivity bounds in the presence of gravity because scattering amplitudes of quantum gravity are less understood in general. If one can derive such bounds, however, the bounds will be understood as the swampland conditions on a given EFT. There are various attempts to formulate positivity bounds in the presence of gravity: see e.g.,~\cite{Hamada:2018dde,Bellazzini:2019xts,Loges:2020trf,Alberte:2020jsk,Tokuda:2020mlf,Herrero-Valea:2020wxz,Alberte:2020bdz,Alberte:2021dnj,Caron-Huot:2021rmr,Caron-Huot:2022ugt,Herrero-Valea:2022lfd,deRham:2022gfe,Noumi:2022wwf,Hamada:2023cyt}.

In this paper, we discuss a particular version of positivity bounds with gravitational effects included \cite{Tokuda:2020mlf}. 
Interestingly, the effect of gravity modifies the positivity bound, and in some cases strengthens it. While the effects of gravity are suppressed by the powers of the Planck scale, gravity can still change the conclusions dramatically since we are interested in small couplings in dark sectors. Indeed, it has been known that if the gravitational positivity bound is applied to the matter-matter scatterings at loop level in $D=4$ dimensions, we obtain nontrivial constrains on the spectrum and renormalizable couplings of light particles well below the Planck scale~\cite{Cheung:2014ega,Andriolo:2018lvp,Chen:2019qvr,Alberte:2020bdz,Aoki:2021ckh,Noumi:2021uuv,Noumi:2022zht}.

The gravitational positivity bound sometimes gives {\it a lower bound} 
on the size of the dark sector couplings.
When combined with upper bounds from observations,
one can then in principle exclude some dark sector models.
The gravitational positivity bound
relies on some assumptions of the UV theory
and hence can be violated in observations. 
This, however, inevitably implies a violation of some well-defined assumptions of the UV theories, and hence
we can derive sharp constraints of UV physics from the experiments of IR physics.

It is an interesting question to systematically apply the gravitational positivity bounds 
to various BSM models. 
It is fair to say, however, that a systematic application of gravitational positivity bounds for 
BSM physics remains mostly unexplored. Moreover, there are important unsolved problems
whose solutions are required for such analysis, as we will see below. 

The goal of this paper is to make the gravitational positivity bound more accessible to a broader audience of 
phenomenologists interested in BSM physics.
In Fig.\,\ref{fig:flow}, we show the procedures to obtain the gravitational positivity bound for scalar fields.
For this purpose, we list a concrete recipe for 
deriving gravitational positivity bounds for a given EFT, and 
list some fine prints in such discussions.
We next illustrate the power and limitations of the bounds in the concrete example of a  massive dark $\mathrm{U}(1)$ gauge boson (dark photon). 
We will find that the resulting bound is impressively strong,
however, for applications for realistic models, there are some theoretical subtleties that need to be addressed.

The rest of this paper is organized as follows. In Sec.~\ref{sec:bound} we summarize the gravitational positivity bound for non-experts.
In Sec.~\ref{sec:recipe} we present a practical recipe for deriving the gravitational positivity bound.
As a concrete illustration of our procedure, we write down the gravitational positivity bound for a massive $\mathrm{U}(1)$ gauge boson in Sec.~\ref{sec:gauge_boson_mass}.
We will find that the resulting bound is very strong and rules out many parameter spaces. One should note, however, that there are important subtleties for realistic model building,
as we will discuss further in Sec.~\ref{sec:toward}. The final section (Sec.~\ref{sec:conclusion}) is devoted to conclusions and discussions.
We included an appendix for gravitational contributions beyond graviton $t$-channel exchange.

\section{Gravitational Positivity Bounds} \label{sec:bound}

In this section we summarize the gravitational positivity bounds of  \cite{Tokuda:2020mlf}.
To make the presentation friendly to non-experts, we outline the basic assumptions, definitions of quantities to be computed, bounds, and their interpretations, and leave their derivations to original references. 
We begin with the non-gravitational case and then include gravity. We will close with some remarks needed for applications to realistic phenomenological model building. 

\subsection{Positivity Bounds without Gravity}\label{subsec:positivity_bound_without_gravity}

Positivity bounds are formulated in terms of scattering amplitudes. For technical simplicity, this subsection focuses on scalar scattering amplitudes in gapped theories.

\paragraph{Assumptions.}

Consider an $s$-$u$ symmetric scattering amplitude $\mathcal{M}(s,t)$ of $AB\to AB$ type in a given low-energy EFT, 
where $s,t,u$ are the standard Mandelstam variables that satisfy $s+t+u=2(m_A^2+m_B^2)$, and $m_A$ and $m_B$ are the masses of the external particles $A$ and $B$, respectively. We assume that the forward amplitude $\mathcal{M}(s,t=0)$ evaluated in the would-be UV complete theory satisfies the following properties:

    \begin{enumerate}
    
    \item Analyticity: The amplitude $\mathcal{M}(s,t=0)$ is analytic on the physical sheet\footnote{There are in general unstable resonances, whose associated poles are however located in other sheets in the complex $s$-plane.} of the complex $s$-plane except for poles and discontinuities on the real axis required by unitarity.
    
    \item Unitarity: The imaginary part is non-negative, i.e., ${\rm Im}\,\mathcal{M}(s,t=0)\geq 0$.
    
    \item $s^2$-boundedness: The amplitude is bounded by $s^2$ at high energy, i.e., $\displaystyle\lim_{|s|\to\infty}\mathcal{M}(s,0)/s^2=0$. This third condition is guaranteed in local quantum field theories thanks to the Froissart bound~\cite{Froissart:1961ux,Martin:1962rt} and the Phragm\'{e}n-Lindel\"{o}f theorem. 
    
    \end{enumerate}

\paragraph{Positivity bounds.}

The aforementioned three properties imply an infinite set of consistency relations among scattering amplitudes evaluated at the UV and IR. Let us define
    \begin{align}
        \label{def_a2n}
        a_{2n}:=\left[\frac{\partial^{2n}\mathcal{M}(s,t=0)}{\partial s^{2n}}\right]_{s=m_A^2+m_B^2}
        \quad
        (n=1,2,\ldots)\,.
        \end{align}
        Then, the following dispersion relation holds from the analyticity and $s^2$-boundedness of the amplitude:
        \begin{align}
        \label{2n-dispersion}
        a_{2n}=\frac{2\cdot(2n)!}{\pi}\int_{m_{\rm th}^2}^\infty ds\,\frac{{\rm Im}\,\mathcal{M}(s,t=0)}{(s-m_A^2-m_B^2)^{2n+1}}
        \quad
        (n=1,2,\ldots)
        \,,
    \end{align}
where $m_{\rm th}$ is the threshold energy, i.e., the mass of the lightest intermediate state. Note that the relation for $n=0$ does not follow under the present assumptions, which requires a stronger assumption that the amplitude is bounded by $s^0$ at high energy. 

The EFT is defined with a UV cutoff which we denote by $\Lambda_{\rm UV}$. The actual cutoff of the EFT, namely the scale of new physics, is unknown from the low-energy perspective. We thus introduce a reference scale $\Lambda$ below which the EFT is {\it assumed} to be valid, i.e., $\Lambda$ is {\it assumed} to satisfy $\Lambda<\Lambda_{\rm UV}$.

\begin{figure}
        \centering
        \includegraphics[width=100mm, bb=0 0 693 454]{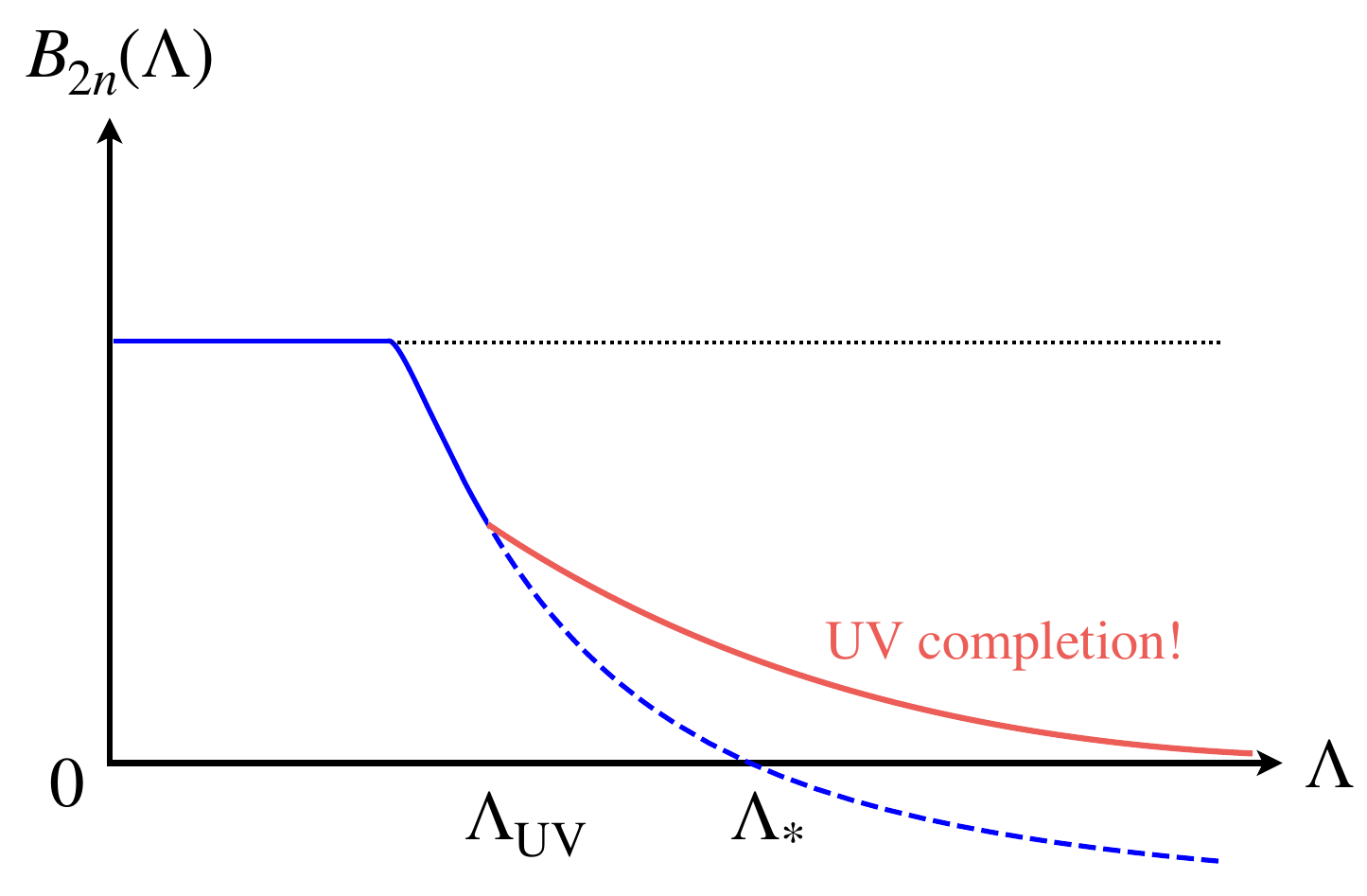}
        \caption{The expression~\eqref{def_B2n_wog} of $B_{2n}(\Lambda)$ is a monotonically non-increasing function as a function of $\Lambda$,
        and crosses zero at the scale $\Lambda=\Lambda_*$.
        For a well-defined UV completion, we need $B_{2n}(\Lambda)\geq0$, which defines an upper bound $\Lambda_*$ on the scale $\Lambda$.}
        \label{fig:improved_positivity}
    \end{figure}

Let us define\footnote{In general, we can have $m_\text{th}^2<m_A^2+m_B^2$ and branch cuts can run on the entire real $s$-axis. The definitions \eqref{def_a2n} and \eqref{def_B2n_wog} may become ill-defined. In such a case, $B_{2n}(\Lambda)$ should be defined through ``arcs'' of~\cite{Bellazzini:2020cot,Bellazzini:2021oaj} (see also~\cite{Arkani-Hamed:2020blm}). $B_{2n}(\Lambda)$ defined in this way agree with \eqref{def_B2n_wog} [or \eqref{defB}] when branch cuts do not run on the entire real $s$-axis, according to the Cauchy integral theorem.}
\cite{Bellazzini:2016xrt,deRham:2017avq,deRham:2017imi}
    \begin{align}
        \label{def_B2n_wog}
        B_{2n}(\Lambda):=a_{2n}
        -\frac{2\cdot(2n)!}{\pi}\int_{m_{\rm th}^2}^{\Lambda^2} ds\,\frac{{\rm Im}\,\mathcal{M}(s,t=0)}{(s-m_A^2-m_B^2)^{2n+1}}
        \quad
        (n=1,2,\ldots)\,.
    \end{align}
We emphasize that $B_{2n}(\Lambda)$ is defined in terms of the amplitude at the IR scale below $\Lambda$, and therefore it is calculable within the EFT. Then, the dispersion relation~\eqref{2n-dispersion} implies
    \begin{align}
        \label{disp_B2n}
        B_{2n}(\Lambda)=\frac{2\cdot(2n)!}{\pi}\int_{\Lambda^2}^\infty ds\,\frac{{\rm Im}\,\mathcal{M}(s,t=0)}{(s-m_A^2-m_B^2)^{2n+1}}\quad
        (n=1,2,\ldots)\,,
    \end{align}
which provides a consistency relation between the IR physics below $\Lambda$ (LHS) and the UV physics above $\Lambda$ (RHS). Furthermore, unitarity implies that the RHS is non-negative, so that
    \begin{align}
        \label{positivity_wog}
        B_{2n}(\Lambda)\geq0
        \quad
        \text{for all $\Lambda$ such that $\Lambda<\Lambda_{\rm UV}$}
        \quad
        (n=1,2,\ldots)\,.
    \end{align}
These bounds are called the positivity bounds. Note that $B_{2n}(\Lambda)$ is a monotonically non-increasing function of $\Lambda$ because the integrand in \eqref{def_B2n_wog} is positive. The typical behavior is shown in Fig.~\ref{fig:improved_positivity}. When the bounds \eqref{positivity_wog} are satisfied at a certain scale, they also hold at any scale below it.
The positivity bound \eqref{positivity_wog} is thus stronger for larger $\Lambda$. 

\paragraph{Interpretations.}
In a given EFT, one can compute the forward amplitude and evaluate $B_{2n}(\Lambda)$ defined in Eq.~\eqref{def_B2n_wog} in terms of parameters of the EFT and a scale $\Lambda$. The positivity bounds~\eqref{positivity_wog} then provide certain inequalities among the EFT couplings and $\Lambda$, leading to two complementary interpretations:

    \begin{enumerate}
    
    \item \underline{Bounds on EFT couplings.} \\ Suppose that the low-energy EFT describes a system below an energy scale $E$. We can then identify $\Lambda$ with the energy scale of interest $E$ to find the bounds on the EFT couplings. These are the necessary conditions for the EFT to have a standard UV completion and be valid below the scale $E=\Lambda$. (In here we are agnostic on the value of the cutoff scale $\Lambda_{\rm UV}$.)
    
    \item \underline{Bounds on the UV cutoff.} \\ In some cases, one may want to assume certain values of the EFT couplings---for example, coupling constants are already fixed by experiments or they need to take certain values for phenomenological purposes (e.g., so that they can be searched in a particular experiment). One can then ask up to which scale the EFT can be valid, i.e., at which scale the new physics comes in. As shown in Fig.~\ref{fig:improved_positivity}, $B_{2n}(\Lambda)$ may become negative at some energy scale $\Lambda_*$, in which case we can derive an upper bound $\Lambda_{\rm UV}<\Lambda_*$ on the UV cutoff of the EFT.
    
    \end{enumerate}

The condition \eqref{positivity_wog} is based on the assumptions on UV, but care must be taken to ensure because we do not know {\it the} UV theory.
If a violation of the inequality \eqref{positivity_wog} is observed in the experiment, it may indicate a violation of the original assumptions.
In this sense, we can also use the positivity bounds \eqref{positivity_wog} to probe UV from IR.

    \begin{figure}
        \centering
        \includegraphics[width=140mm, bb=0 0 863 175]{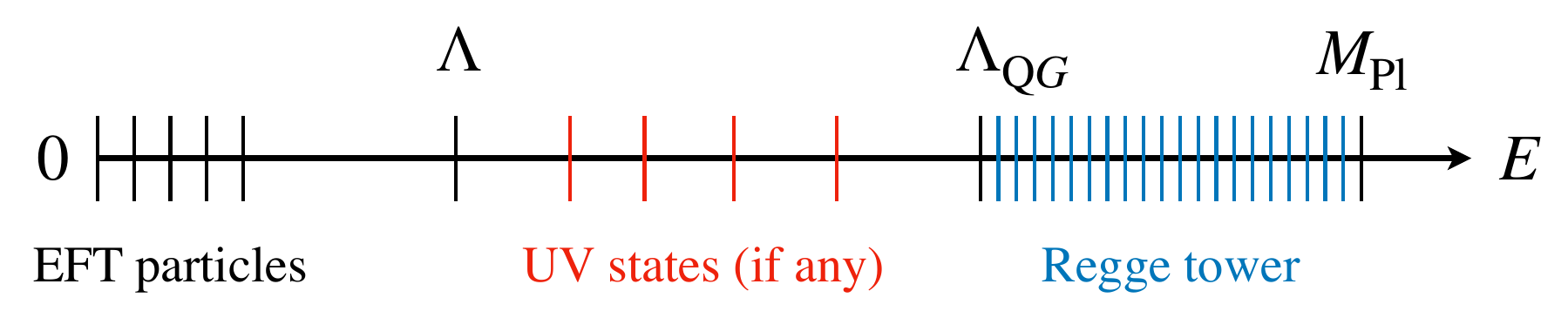}
        \caption{Typical scenario: a UV completion of gravity is achieved well below the Planck scale by an infinite tower of higher-spin states (Regge tower). There may also exist new UV states beyond the EFT that are described within QFT model building and that are not directly related to the UV completion of gravity.}
        \label{fig:scales}
    \end{figure}

\subsection{Positivity Bounds with Gravity}\label{subsection:g_bounds}

We next explain how the positivity bounds are extended to theories with gravity. 
The main new ingredient is that the $t$-channel graviton exchange contributes a new singularity to the low-energy scattering amplitudes as $\sim -s^2/(M_{\rm Pl}^2t)$ ($M_{\rm Pl}\simeq 2.4\times 10^{18}\, \textrm{GeV}$ being the reduced Planck mass), invalidating the definition~\eqref{def_a2n} of $a_2$ and hence the positivity argument for $B_2(\Lambda)$.\footnote{Note that the positivity argument for $B_{2n}$ ($n=2,3,\ldots$) is unchanged. In the following, we focus on $B_2$.} In this subsection, we summarize how the positivity bounds are formulated in the presence of gravity.

\paragraph{Assumptions.}
The basic assumptions are the same as before, namely analyticity, unitarity, and the $s^2$-boundedness $\displaystyle\lim_{|s|\to\infty}\bigl|\mathcal{M}(s,t)/s^2\bigr|=0$ for small negative $t$. 
\\

While the argument below can be applied to any UV complete theory with these three properties, a typical candidate of such a UV completion is a perturbative string theory with a particle spectrum as shown in Fig.~\ref{fig:scales}.

\paragraph{Gravitational positivity bounds.}

We define $a_2$ by subtracting the graviton $t$-pole from the amplitude and taking the forward limit as\footnote{
For simplicity, we focus on the amplitude up to $\mathcal{O}(\Mpl^{-2})$ and only subtract the graviton $t$-pole as a divergent term in the forward limit $t \to -0$ in the definition of $a_2$. See \cite{Herrero-Valea:2020wxz, Herrero-Valea:2022lfd} for arguments with the $\mathcal{O}(M_{\rm Pl}^{-4})$ terms.}
    \begin{align}
        \label{def_a2}
        a_2 &:= \lim_{t\to -0} \left[ \frac{\partial^2 \mathcal{M}(s,t) }{\partial s^2}  + 
        \frac{2}{M_{\rm Pl}^2 t}
        \right]_{s=m_A^2+m_B^2}
        \,.
    \end{align}
With $a_2$ defined in this manner, we introduce
    \begin{align}
        B(\Lambda)&:=a_2-\frac{4}{\pi}\int_{m_{\rm th}^2}^{\Lambda^2} ds\,\frac{{\rm Im}\,\mathcal{M}(s,t=0)}{\left(s-m_A^2-m_B^2\right)^3}\,.
        \label{defB}
    \end{align}
Here and in what follows we suppress the subscript $2$ of $B_2(\Lambda)$ for notational simplicity. The dispersion relation~\eqref{disp_B2n} is then modified as
    \begin{align}
        \label{B_Lambda}
        B(\Lambda)=\lim_{t\to-0}\left[
        \frac{2}{M_{\rm Pl}^2 t}
        +\frac{4}{\pi}\int_{\Lambda^2}^\infty ds\,\frac{{\rm Im}\,\mathcal{M}(s,t)}{\left(s+\frac{t}{2}-m_A^2-m_B^2\right)^3}\right]\,.
    \end{align}
Due to the first term in the square bracket (which is negative for $t<0$), the sign of $B(\Lambda)$ is undetermined. 
However, one can show that the positivity of $B(\Lambda)$ holds at least approximately. The idea is that we split the integral of the dispersion relation into the quantum gravity part and the other,
    \begin{align}
        \label{B_LambdaQG}
        B(\Lambda)=\frac{4}{\pi}\int_{\Lambda^2}^{\Lambda_{\rm QG}^2} ds\,\frac{{\rm Im}\,\mathcal{M}(s,0)}{\left(s-m_A^2-m_B^2\right)^3}+
        \lim_{t\to-0}\left[
        \frac{2}{M_{\rm Pl}^2 t}
        +\frac{4}{\pi}\int_{\Lambda_{\rm QG}^2}^{\infty} ds\,\frac{{\rm Im}\,\mathcal{M}(s,t)}{\left(s+\frac{t}{2}-m_A^2-m_B^2\right)^3}\right] \,.
    \end{align}
When we take the forward limit $t\to-0$, the l.h.s of (\ref{B_LambdaQG}) is finite because it is defined by subtracting the $t$-pole term. 
Then, although the $t$-pole term in the r.h.s diverges in the forward limit $t\to -0$, 
the sum in the square bracket should be finite as a result of a cancellation with the quantum gravity part:
    \begin{align}\label{M_def}
        \lim_{t\to-0}\left[
        \frac{2}{M_{\rm Pl}^2 t}
        +\frac{4}{\pi}\int_{\Lambda_{\rm QG}^2}^{\infty} ds\frac{{\rm Im}\,\mathcal{M}(s,t)}{\left(s+\frac{t}{2}-m_A^2-m_B^2\right)^3}\right]
        =: \frac{\mysigma}{M_{\rm Pl}^2M^2}
        \quad
        (\mysigma=0, \pm 1)
        \,,
    \end{align}
where we defined the sign $\mysigma$ and the mass scale $M$, which depend on details of the cancellation mechanism. {\color{black} (The case $\sigma=0$ corresponds to the exact cancellation.)}
As a result, we find
    \begin{align}
        \label{g-positivity}
        B(\Lambda)\geq\frac{\mysigma}{M_{\rm Pl}^2M^2}
         \quad
        \text{for all $\Lambda$ s.t.\
        $\Lambda<\Lambda_{\rm UV}< \Lambda_{\rm QG}$
        } \, ,
    \end{align}
which we call the {\it gravitational positivity bound}.

The right-hand side of \eqref{g-positivity} is suppressed by powers of $\Mpl$, and hence
disappears in the decoupling limit $M_{\rm Pl}\to \infty$, where the gravitational positivity bound reduces to the 
positivity bound without gravity. 

While the derivation here works irrespective of the details of the cancellation mechanism, 
the most plausible scenario is the Regge behavior\footnote{See also \cite{Camanho:2014apa,DAppollonio:2015fly} for connections between Reggeization and causality.}: the gravitational amplitude is modified as ${\rm Im}\mathcal{M} \sim s^{2+\alpha't+\cdots}$ with $\alpha' \sim \Lambda_{\rm QG}^{-2}$ by an infinite tower of higher-spin states at $s \ge \Lambda_\text{QG}^2$ (see Fig.~\ref{fig:scales}).
The perturbative string theory is a typical example of this scenario in which the quantum gravity scale $\Lambda_\text{QG}$ corresponds to the string scale. The cancellation of the $t$-pole can be explicitly shown under the assumption of Regge behavior~\cite{Tokuda:2020mlf}.

\paragraph{Interpretations.}

In contrast to the non-gravitational positivity bound \eqref{positivity_wog}, the gravitational positivity bound~\eqref{g-positivity} contains not only the IR data $B(\Lambda)$ determined by the EFT parameters up to the scale $\Lambda$, but also the UV data $(M, \mysigma)$ of gravitational Regge amplitudes (or, generically speaking, quantum gravity amplitudes). Therefore, there are three possible interpretations, depending on the context~\cite{Noumi:2021uuv, Noumi:2022zht, Alberte:2020bdz, Aoki:2021ckh, Alberte:2021dnj}:

    \begin{enumerate}
    
    \item \underline{Quantum gravity constraints on IR physics.} \label{interpretation1}
    
    If we specify/assume a quantum gravity scenario and the parameters $(\mysigma,M)$ of gravitational Regge amplitudes, we can think of the bounds~\eqref{g-positivity} as quantum gravity constraints on the low-energy EFT at a scale $\Lambda=E$. Such constraints are useful to carve out the parameter space of phenomenological models in the spirit of the Swampland Program\footnote{\textcolor{black}{See \cite{Alberte:2020bdz, Cheung:2014ega, Hamada:2018dde,Andriolo:2018lvp} for possible connections between positivity and the Swampland conjecture.}}.
    
    \item \underline{Constraints on the scale of new physics.} \label{interpretation2}
    
    As in the non-gravitational case, we can interpret \eqref{g-positivity} as the bounds giving the maximum cutoff $\Lambda_*$ when we assume the parameters of the EFT and the UV data $(\mysigma, M)$. When the bounds \eqref{g-positivity} are violated, one can try to increase the cutoff of the EFT by adding new state(s) to EFT like Fig.~\ref{fig:improved_positivity}. If the bounds \eqref{g-positivity} are inevitably violated at $\Lambda>\Lambda_*$, this is a sign of the necessity for a UV completion of gravity. In this case, $\Lambda_*$ is not a maximum scale of just new physics, but the upper bound on the quantum gravity scale $\Lambda_{\rm QG}$. 
    
    \item \underline{IR constraints on quantum gravity.} \label{interpretation3}
    
    Once the parameters and the validity of the model are identified by experiments at a scale $\Lambda=E$, we can use the bounds~\eqref{g-positivity} to constrain the parameters $(\mysigma,M)$ of the gravitational Regge amplitudes required for UV completion of gravity. 
    Such constraints are useful as a necessary condition for a quantum gravity theory to describe our real world.
    
    \end{enumerate}

As we mentioned in Sec.~\ref{subsec:positivity_bound_without_gravity}, the violation of original assumptions on UV theory leads to the violation of \eqref{g-positivity}.
Differently from the non-gravitational case, however, experimental tests of the violation of \eqref{g-positivity} require knowledge of the UV data $(M,\sigma)$.

\paragraph{\textcolor{black}{Comments on $(M,\sigma)$.}}

\textcolor{black}{
As we explain more explicitly later, nontrivial constraints on the IR physics are obtained in the spirit of Interpretation \ref{interpretation1} when $B_{\rm grav}-\sigma/(\Mpl^2 M^2)<0$. Here, $B_{\rm grav}$ is the gravitational part of $B$ which will be defined in the next section. This condition is satisfied either when $\sigma=0, +1$ or when 
$M$ is sufficiently large, typically 
$M\gg m_\text{light}$ with $m_\text{light}$ being the mass scale of light particles. This shows that details of the UV data $(M,\sigma)$ are important in discussing phenomenological implications of the gravitational positivity~\cite{Noumi:2022zht}.
}


\textcolor{black}{
Notably, recent studies proceed in the direction to carve out the parameter space of $(M,\sigma)$ itself and sharpen the gravitational positivity bound~\eqref{g-positivity}. For scattering of identical scalars, the best bound obtained so far is schematically of the form~\cite{Caron-Huot:2021rmr,Noumi:2022wwf}\footnote{
\textcolor{black}{
There are two complementary approaches to this problem depending on how to treat the graviton $t$-channel pole. First, the one initiated in~\cite{Caron-Huot:2021rmr} works in finite impact parameter to make the graviton contribution finite. While it gives a universal bound that does not rely on details of UV completion, the obtained bound is trivialized in spacetime four dimensions due to a logarithmic IR divergence. On the other hand, the one proposed in~\cite{Noumi:2022wwf} uses the sum rule of gravitational Regge amplitudes to give a bound on the parameter space $(M,\sigma)$. While this approach assumes Reggeization of gravitational amplitudes, it is applicable even in spacetime four dimensions.
}
} (see also~\cite{deRham:2022gfe,Hamada:2023cyt}),
\begin{align}
    B(\Lambda)
    \geq 
    \frac{-\mathcal{O}(1)}{\Mpl^2 m^2}
    \,,
\end{align}
where $m$ denotes the threshold energy, i.e., the energy of the lightest intermediate state.
For tree-level amplitudes, this scale is essentially the mass of the lightest higher-spin particle (spin $2$ or higher). For loop amplitudes, this corresponds to the lowest energy of the intermediate multi-particle states. Note that this bound is a necessary condition for the effective theory to have a standard UV completion, rather than a sufficient condition. Therefore, it is still a nontrivial question if there exists a consistent quantum gravity theory that accommodates a negative $B(\Lambda)$,\footnote{
\textcolor{black}{
To our best knowledge, there is no known tree-level string amplitude that accommodates $\sigma=-1$, even though $\sigma=-1$ is not prohibited by the consistency of scalar scattering alone. It would be interesting to clarify in what class of quantum gravity theories (if exist) $\sigma=-1$ can be realized.
}
}
more specifically that with $\sigma=-1$ and $M\sim m_{\rm light}$. Further studies in this direction are encouraged for phenomenological applications of the gravitational positivity.
}

\subsection{Generalizations for Realistic Theories} \label{sec:realistic}

When one tries to apply the gravitational positivity bounds \eqref{positivity_wog} to realistic phenomenological models, 
the basic idea is the same as in the case of scalar scatterings of the previous subsection.
However, one encounters several new subtleties to be carefully addressed, which we list in the following.

\paragraph{Spin.} 
The positivity bound can be extended to a scattering of particles with spin, see~\cite{Bellazzini:2016xrt,deRham:2017zjm}.\footnote{
For this purpose, it is useful to diagonalize the crossing relation using the transversity formalism of \cite{Kotanski:1970}. For the use of this formalism in the context of positivity bounds, see \cite{deRham:2017zjm}.}
There are extra technical complications, however, and here we highlight the issue of the kinematic singularity, 
which is relevant for our discussion of gauge bosons in Sec.~\ref{sec:recipe}.

Let us consider two-to-two scattering amplitudes of massive spin-1 particles with mass $m_V$. The kinematic singularity is the singularity of the scattering amplitude $\mathcal{M}$ from the pole $s=4 m_V^2$ of the 
 scattering angle $\theta$ as in
    \begin{align}
        \cos\theta = 1 + \frac{2t}{s-4m_V^2}\ .
    \end{align}
The kinematic singularity disappears in the forward limit $t \to 0$ if the scattering amplitude does not contain the massless $t$-pole. 
However, in our setup, the kinematic singularity appears from $t$-channel graviton exchange diagrams.
We can deal with this singularity by defining 
a kinematic singularity-free amplitude to be \cite{deRham:2017zjm} (see also \cite{Wang:1966zza,Cohen-Tannoudji:1968lnm}):\footnote{\textcolor{black}{
Although the $s^2$ factor is not necessary to cancel the kinematic singularity, it simplifies the subsequent expressions by retaining $s \leftrightarrow u$ crossing symmetry in $t \to 0$ limit.
}}
    \begin{align}
        \label{tilde_M}
            \widetilde{\mathcal{M}}(s,t) := 
            s^2
            \left( s-4m_V^2 \right)^2 \mathcal{M}(s,t) \ .
    \end{align}
We can then derive the positivity bound by essentially the same arguments, as long as we replace the amplitude $\mathcal{M}$ by $ \widetilde{\mathcal{M}}$.
In practice, this is the same as ignoring the contribution of kinematic singularity to the $t$-channel graviton exchange diagrams.

\paragraph{Massless particles.} The general properties of the scattering amplitudes have been established only in gapped systems, and not when massless particles are present. 
Even worse, the traditional non-perturbative S-matrix does not exist for massless particles, due to the issue of the IR divergence. 
To avoid this IR issue, we estimate the contributions from IR divergent diagrams to the positivity bounds by introducing an IR cutoff 
\textcolor{black}{in Appendix~\ref{sec:non_grav}.}
We find that they will be subdominant, and hence we expect that the subtleties associated with the IR divergence will not affect our main conclusion. 
However, this may not be the case in other models, in which case an appropriate prescription for massless particles is required to derive the bounds.

\paragraph{Unstable particles.} If a particle has a decay channel, such an unstable particle does not appear in the asymptotic states and the standard scattering amplitudes cease to exist for the particle~\cite{Veltman:1963th}. However, unstable-particle scattering amplitudes can be unambiguously defined by residues of higher-point amplitudes (see e.g.~\cite{Eden:1966dnq}) and their properties can be studied in the 
S-matrix theory.\footnote{There are still IR divergences, which need to be addressed separately.} Unitarity of the S-matrix leads to certain constraints on the unstable-particle scattering amplitudes, and in particular, there exists a positivity constraint on the imaginary part~\cite{Aoki:2022qbf}, suggesting the existence of positivity bounds even from scattering of unstable particles. In the present paper, however, we will only focus on the scattering of stable particles and leave a precise treatment of unstable particles for a future study.

\paragraph{Anomalous threshold singularities.} 
Even for stable particles, the analytic structure of a scattering of a heavier particle is not as simple as that of a lighter particle. 
It is known that the scattering of the heavier particle can have a new singularity called an anomalous threshold, whose existence does not immediately follow from unitarity. While general knowledge of anomalous threshold is still missing to the best of our knowledge (see~\cite{Hannesdottir:2022bmo, Correia:2022dcu} for recent discussions), 
we expect that the anomalous thresholds appear only in the IR regime, i.e., they are detectable and controlled within EFT at least when the heavy particle is stable. If this is indeed the case, the anomalous thresholds will give additional technical complications but will not spoil the success of positivity bounds, as long as these thresholds are taken into account in the derivation of the dispersion relations.

\section{Practical Recipe for Gravitational Positivity}\label{sec:recipe}

We are now ready to discuss practical recipes for deriving gravitational positivity bounds for EFTs.
Instead of making the presentation fully general,
we focus on the case of $2\to 2$ scattering $VV\to VV$ of spin $1$ 
particle $V$ with mass $m_V$.
This will be the case relevant for the rest of this paper.

The gravitational positivity bound is an  inequality for the quantity $B(\Lambda)$. When we wish to apply the gravitational positivity bound of the previous section, 
we run into one practical problem:
we can almost never compute an exact scattering amplitude $\mathcal{M}(s,t)$ in an EFT. 
In fact, if we know the exact expression for $\mathcal{M}(s,t)$, we can analytically continue the expression to the whole complex plane, and we have effectively solved for the quantum gravity already. The best we can do is to 
compute an approximate expression $\mathcal{M}_{\rm EFT}(s,t)$ for $\mathcal{M}(s,t)$, by including only a finite number of parameters out of an infinite number of 
higher-dimensional operators. 
The differences between the two are small in the regions of the validity of EFT (at scales $\Lambda \ll \Lambda_{\rm UV}$). Hence, the computation of the {\it approximate amplitude} 
$\mathcal{M}_{\rm EFT}$ is sufficient for evaluating the quantity $B(\Lambda)$ which only requires information below $\Lambda$. In the following, we will discuss EFT scattering amplitude $\mathcal{M}_{\rm EFT}$ and explain a practical way to calculate $B(\Lambda)$ from the EFT amplitude $\mathcal{M}_{\rm EFT}$: for notational simplicity we will simply denote this approximate amplitude as $\mathcal{M}$ in the rest of this section.

We consider the scattering amplitude of the gauge bosons $VV \to VV$. 
Corresponding to polarizations of the gauge boson, we define the following combinations of helicity amplitudes. 
For the scattering of transverse modes,
\begin{align}\label{M_TT}
    \mathcal{M}_{TT}(s, t):= & \frac{1}{4}\left[\mathcal{M}\left(1^{+} 2^{+} 3^{+} 4^{+}\right)+\mathcal{M}\left(1^{+} 2^{-} 3^{+} 4^{-}\right)+\mathcal{M}\left(1^{-} 2^{-} 3^{-} 4^{-}\right)+\mathcal{M}\left(1^{-} 2^{+} 3^{-} 4^{+}\right)\right]\ ,
\end{align}
where superscripts $\pm$ denote helicities.
For the scattering of the transverse mode and the longitudinal mode,
\begin{align}\label{M_TL}
    \mathcal{M}_{TL}(s, t):=\frac{1}{2}\left[\mathcal{M}\left(1^{+} 2^L 3^{+} 4^L\right)+\mathcal{M}\left(1^{-} 2^L 3^{-} 4^L\right)\right]\ ,
\end{align}
where superscripts $L$ denote longitudinal polarizations.
For the scattering of longitudinal modes,
\begin{align}\label{M_LL}
    \mathcal{M}_{LL}(s, t):=\mathcal{M}\left(1^{L} 2^L 3^{L} 4^L\right)\ .
\end{align}
These amplitudes are defined such that they exhibit $s \leftrightarrow u$ symmetry (up to subtleties associated with the kinematical singularity discussed around \eqref{tilde_M}, which are not relevant to the following discussion).

 We provide a step-by-step procedure for deriving the gravitational positivity bound. The basic task is simple: we need to 
evaluate the cutoff-dependent quantity $B(\Lambda)$ defined by
a spin-$1$ generalization of \eqref{B_Lambda}:
\begin{align}
B(\Lambda):=a_2-\frac{4}{\pi}\int_{m_{\rm th}^2}^{\Lambda^2} ds\,\frac{{\rm Im}\, \widetilde{\mathcal{M}}(s,t=0)}{\left(s-2 m_V^2\right)^7}\,,
\label{defB2}
\end{align}
where $\widetilde{\mathcal{M}}$ is a kinematic singularity-free amplitude defined in \eqref{tilde_M}, $m_{\rm th}$ is the threshold energy, i.e., the mass of the lightest intermediate state, and $a_2$ is basically defined as in \eqref{def_a2}, but with gravitational corrections included (as we will soon see in \eqref{B_a2}).

\bigskip
\noindent 
{\bf Step 1: Calculate $\mathcal{M}(s,t)$ }

We compute the $2\to 2$ scattering amplitude $\mathcal{M}(s,t)$
up to order $\mathcal{O}\left(M_{\text{Pl}}^{-2}\right)$:
\begin{align}\label{eq:M_as_sum}
    \mathcal{M}(s,t) = \underbrace{\mathcal{M}_{\text{non-grav}}(s,t)}_{\mathcal{O}\left(M_{\text{Pl}}^{0}\right)} + \underbrace{\mathcal{M}_{\text{grav}}(s,t)}_{\mathcal{O}(M_{\text{Pl}}^{-2})} + \underbrace{ \vphantom{\mathcal{M}_{\text{grav}}} \cdots  }_{\mathcal{O}(M_{\text{Pl}}^{-4})}\ .
\end{align}
Here $\mathcal{M}_{\text{non-grav}}(s,t)$ is the contribution from diagrams with no graviton exchange, while $\mathcal{M}_{\text{grav}}(s,t)$ is the contribution from diagrams with one graviton exchange and is of the order $\mathcal{O}\left(M_{\text{Pl}}^{-2}\right)$. The remaining terms are the order of $\mathcal{O}\left(M_{\text{Pl}}^{-4}\right)$ and are neglected in the rest of the analysis. 
Once we compute $\mathcal{M}$, it is straightforward to obtain
kinematic-singularity free amplitude $\widetilde{\mathcal{M}}$ \eqref{tilde_M},
both for non-gravitational and gravitational contributions.
Note that such an amplitude exists for each choice of helicity of external vector fields (as in \eqref{M_TT}, \eqref{M_TL}, and \eqref{M_LL}).

\bigskip
\noindent 
{\bf Step 2: Calculate $B_{\text{non-grav}}\left( \Lambda \right)$}

We compute the non-gravitational contribution $B_{\text{non-grav}}(\Lambda)$ to $B(\Lambda)$.
We assume that the EFT of interest is renormalizable in the gravity decoupling limit $M_{\rm Pl}\to \infty$.

In such a case, the amplitude $\mathcal{M}_{\text{non-grav}}$ itself satisfies the analyticity and the $s^2$-boundedness, which is the $s^6$-boundedness in terms of $\mathcal{\widetilde{M}}_{\text{non-grav}}$, when the expression is analytically continued throughout the complex plane, even outside the regions of validity of the EFT.
The non-gravitational part then satisfies the twice-subtracted dispersion relation, giving a formula
    \begin{align}
    B_{\text{non-grav}}(\Lambda) =\frac{4}{\pi}\int_{\Lambda^2}^{\infty} ds\, \frac{{\rm Im}\,\widetilde{\mathcal{M}}_{\text{non-grav}}(s,t=0)}{\left(s-2m_V^2\right)^7}\ .
    \end{align}
This formula is practically useful since we only need to compute the imaginary part of the forward amplitude.
This expression further simplifies in the limit $\Lambda \gg m_V$:
\begin{screen}
    \begin{align}
    \label{B_nongrav_approx}
    B_{\text{non-grav}}(\Lambda) \simeq\frac{4}{\pi}\int_{\Lambda^2}^{\infty} ds\, \frac{{\rm Im}\,\mathcal{M}_{\text{non-grav}}(s,t=0)}{\left(s-2m_V^2\right)^3} 
    \;.
    \end{align}
\end{screen}
In other words, in this limit, we can in practice neglect the subtleties associated with the use of the kinematic-singularity-free amplitude 
$\widetilde{\mathcal{M}}$, and use the original amplitude $\mathcal{M}$ for the evaluation of $B_{\text{non-grav}}(\Lambda)$ in \eqref{B_nongrav_approx} (which coincides \eqref{disp_B2n} with $n=1$ and $m_A=m_B=m_V$).

\bigskip
\noindent 
{\bf Step 3: Calculate $B_{\text{grav}}\left( \Lambda \right)$}

We next compute the gravitational contribution $B_{\text{grav}}(\Lambda)$ to $B(\Lambda)$. While this contains many different Feynman diagrams in general,
we will argue in Appendix \ref{sec:non_grav} that
it is enough to consider the contribution from $t$-channel graviton exchange diagrams.
In this case, the integral part of (\ref{defB}) is absent for this contribution because $\widetilde{\mathcal{M}}_{\text{grav,$t$-channel}}(s,t)$ does not have the $s$-channel or $u$-channel discontinuity at the non-gravitational one-loop order. 
Therefore the gravitational part of $B\left( \Lambda \right)$ is simplified to $a_2$ defined in \eqref{def_a2}, evaluated 
for the gravity $t$-channel amplitude:
    \begin{align}
        \label{B_a2}
        B_{\text{grav}}\left( \Lambda \right) &\simeq a_{2,\text{grav,$t$-channel}} :=
        \lim_{t\to -0} \left[ \frac{2}{6!} \frac{\partial^6 \mathcal{\widetilde{M}_{\text{grav,$t$-channel}}}(s,t) }{\partial s^6}  + 
        \frac{2}{M_{\rm Pl}^2 t} 
        \right]_{s=2 m_V^2}
     \, .
    \end{align}
When we assume $\Lambda \gg m_V$, we obtain an 
expression
    \begin{screen}
        \begin{align}
        B_{\text{grav}}\left( \Lambda \right) \,\simeq \lim_{t\to -0} \left[ \frac{\partial^2 \mathcal{M_{\text{grav,$t$-channel}}}(s,t) }{\partial s^2}  + 
        \frac{2}{M_{\rm Pl}^2 t} 
        -\text{(kinematic singularity)} \right]_{s=2 m_V^2}\, .
        \end{align}  
    \end{screen}

\bigskip
\noindent 
{\bf Step 4: Write down the inequality for total $B(\Lambda)$}

The constraints on EFT are obtained as
the inequality \eqref{g-positivity}
for the combined expression
$B(\Lambda) = B_{\text{non-grav}}(\Lambda) + B_{\text{grav}}(\Lambda)$.
There are three interpretations of this inequality, as we discussed in Sec.~\ref{subsection:g_bounds}.

\section{\texorpdfstring{Positivity Bound on Dark U(1) Gauge Boson}{Positivity Bound on Dark U(1) Gauge Boson}}
\label{sec:gauge_boson_mass}

A gauge interaction is one of the cornerstones of the quantum field theory (QFT).
Gauge symmetries and their breakings are one of the most crucial ingredients of the 
Standard Model, and many BSM models also introduce new gauge symmetries.
In particular, a light gauge boson is one of the candidates for the DM~\cite{Nelson:2011sf,Arias:2012az,Graham:2015rva}.

In this section, we work out the details of the gravitational positivity recipe for the dark $\mathrm{U}(1)$ gauge boson.
Here we study the Abelian Higgs mechanism of the simplest gauge theory, the $\mathrm{U}(1)$ theory. 
We discuss theoretical constraints obtained by step-by-step computations spelled out in the previous section.

\subsection{Higgs Contribution}

In the Higgs mechanism, the gauge boson gets a mass when a charged scalar field $\Phi$ gets a non-zero vacuum expectation value (VEV) $v$. 
The renormalizable Lagrangian of the Higgs mechanism to generate a mass for a $\mathrm{U}(1)$ gauge boson is 
    \begin{align}\label{def:L}
        \mathcal{L} = |D_{\mu} \Phi|^2 - \frac{\lambda}{4} (|\Phi|^2  - v^2)^2 \, ,
    \end{align}
where $D_\mu = \partial_\mu -  i g_\Phi  V_\mu$ is the covariant derivative and $g_\Phi$ is a charge of $\Phi$ and $\lambda > 0$.
After $\Phi = v + \phi/\sqrt{2} + i G$ develops the VEV, the Goldstone component $G$ is absorbed into the longitudinal component of the gauge boson.
There are a massive gauge boson $V$ and a real scalar $\phi$ after the gauge symmetry breaking, 
and the gauge boson gets a mass $m_V = \sqrt{2} g_\Phi v$ and a real scalar $\phi$ has a mass $m_\phi = \sqrt{\lambda} v$.
The interaction of the Higgs and gauge bosons are given by
    \begin{align}
        \mathcal{L} \ni \left(\frac{g_\Phi^2}{2} \phi^2 + g_\Phi m_{V} \phi  \right) V^{\mu} V_{\mu}
         -\lambda \left(\frac{v }{2\sqrt{2}} \phi^3 + \frac{1}{16} \phi^4\right) \, \label{eq:higgs_int}.
    \end{align}
The gravitational interaction can be found by expanding the Lagrangian (\ref{def:L}) in terms of the canonical gravitational field $h_{\mu\nu}$ as
    \begin{align}
        {\cal L}_{\rm grav} = -\frac{h_{\mu\nu}T^{\mu\nu}}{\Mpl} + {\cal O}(\Mpl^{-2}) \,; \quad T^{\mu\nu} \equiv -\frac{2}{\sqrt{-g}}\frac{\partial (\sqrt{-g}{\cal L})}{{\partial g_{\mu\nu}}} \,.
    \end{align}
The Higgs-gauge boson interaction \eqref{eq:higgs_int} does not make the gauge boson $V$ unstable since it has a $\mathbb{Z}_2$ symmetry $V \leftrightarrow -V$.

\subsubsection{Calculation of $B(\Lambda)$}

    \begin{figure}[t!]
    	\centering
     	\subcaptionbox{Non-gravitational Feynman diagrams.\label{fig:nongra_h}}
    	{\includegraphics{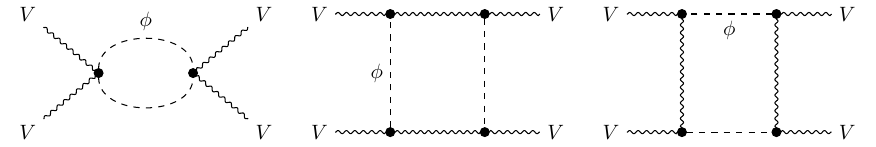}}
     	\subcaptionbox{
      $t$-channel graviton exchange diagrams.\label{fig:gra_h}}
    	{\includegraphics[trim=0 0 0 -20]{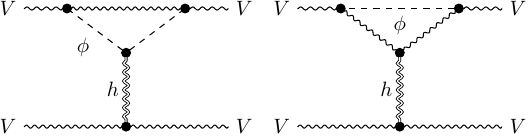}}
     \caption{Some examples of the Higgs loop contribution to $B(\Lambda)$.}
     \label{fig:Higgs-loop2}
    \end{figure}	

We calculate $B\left( \Lambda \right)$ at the one-loop order using the recipe described in Sec.~\ref{sec:recipe}.
In Figs.\,\ref{fig:nongra_h} and \ref{fig:gra_h}, we show some examples of the gravitational and non-gravitational processes.
The one-loop diagrams are calculated by using the Mathematica packages FeynRules~\cite{Christensen:2009jx,Christensen:2008py,Alloul:2013bka}, FeynArts~\cite{Hahn:2000kx},
FeynCalc~\cite{Mertig:1990an,Shtabovenko:2020gxv,Shtabovenko:2016sxi} and Package-X~\cite{Patel:2015tea}.

\subsubsection*{Non-gravitational contributions}
Fig.\,\ref{fig:nongra_h} show some examples of the non-gravitational Feynman diagrams involving the Higgs boson $\phi$.
From these (and other) diagrams we evaluate the non-gravitational part of $B(\Lambda)$ to be
{\footnotesize
\begin{align}
    B^{TT}_{\text{non-grav}}&= \frac{g_\Phi^4}{4 \pi ^2 \Lambda ^4} + \frac{g_{\Phi }^4 \left[ m_\phi^4 ( 12 \log \left(\frac{m_\phi}{\Lambda }\right)+1)+ m_\phi^2 m_V^2(12 \log
    \left(\frac{\Lambda ^3 m_V}{m_\phi^4}\right)-27)+24 m_V^4\right]}{18 \pi ^2 \Lambda ^6
    m_\phi^2} + \cdots \,,
    \label{BTTng} \\
     B^{TL}_{\text{non-grav}}&= \frac{g_\Phi^4}{2 \pi ^2 \Lambda ^4}+
    \frac{g_{\Phi }^4 \left[-3 \left(m_\phi^3-2 m_\phi m_V^2\right){}^2 \log \left(\frac{m_\phi m_V}{\Lambda ^2}\right)+17
   m_\phi^4 m_V^2-20 m_\phi^2 m_V^4-8 m_\phi^6+12 m_V^6\right]}{9 \pi ^2 \Lambda ^6 m_\phi^2 m_V^2} + \cdots \,, \label{BTLng} \\
       B^{LL}_{\text{non-grav}}&=\frac{g_\Phi^4}{\pi ^2 \Lambda ^2 m_{V}^2} + \frac{g_{\Phi }^4 \left(4 m_\phi^2 m_V^2 \log \left(\frac{m_V^2}{\Lambda ^2}\right)-8 m_\phi^2 m_V^2+5 m_\phi^4+6
    m_V^4\right)}{8 \pi ^2 \Lambda ^4 m_V^4} +\cdots \,.  \label{BLLng}
\end{align}
}
Here $\cdots$ represents contributions suppressed by higher powers of $\Lambda$, which scale is assumed to be  much greater than $m_\phi$ and $m_V$.
We have also numerically estimated the two-loop contributions and confirm that these effects are smaller than the one-loop contributions in the parameter region of present interest.

\subsubsection*{Gravitational contributions}
\begin{figure}[t]
    \centering
    \includegraphics[scale=0.8]{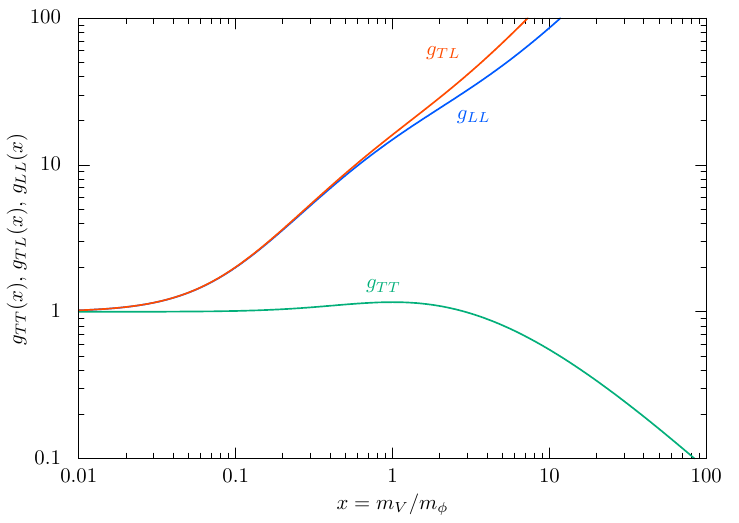}
    \caption{The functions $g_{{TT, TL, LL}}(x)$.
    These functions are real and positive for $x>0$ and therefore the gravitational contributions are negative for all the masses of the Higgs boson and the gauge boson.
    }
    \label{fig:gfun}
\end{figure}

As the next step, we compute the one-loop diagrams of the gravitational contributions (Fig.\,\ref{fig:gra_h}):
\begin{align}
B^{TT}_{\mathrm{grav}} &= 
\frac{-g_\Phi^2}{72 \pi^2 m_\phi^2 \Mpl^2}
\, g_{{TT}}\!\left( \frac{m_V}{m_{\phi}} \right) \, , \\
B^{TL}_{\mathrm{grav}} &= 
\frac{-g_\Phi^2}{144 \pi^2 m_V^2 \Mpl^2}
\, g_{{TL}}\!\left( \frac{m_V}{m_{\phi}} \right) \, ,  \\
B^{LL}_{\mathrm{grav}} &= 
\frac{-g_\Phi^2}{72 \pi^2 m_V^2 \Mpl^2}
\, g_{{LL}}\!\left( \frac{m_V}{m_{\phi}} \right) \, ,
\end{align}
where the functions $g_{TT, TL, LL}$ are given by
{\footnotesize
\begin{align}
g_{{TT}}(x)&= \frac{2}{x^8 \left(1-4 x^2\right)^2} \cr
&\times \Bigg[-6 \left(2 x^2-1\right) \left(4 x^4-5
   x^2+1\right)^2 \log (x)+\left(200 x^8-346 x^6+230 x^4-63
   x^2+6\right) x^2 
   \cr 
   & -6 \sqrt{\frac{1}{x^2}-4} \left(4 x^{10}-32
   x^8+52 x^6-35 x^4+10 x^2-1\right) x \log \left(\frac{1}{2}
   \left(\sqrt{\frac{1}{x^2}-4}+\frac{1}{x}\right)\right)
   \Bigg]
   \, ,\\
 g_{{TL}}(x) &=  \frac{-1}{x^6 \left(1-4 x^2\right)^2} \cr
& \times \Bigg[6 \sqrt{\frac{1}{x^2}-4} x \left(51 x^2+2 \left(8 x^6-92
   x^4+147 x^2-93\right) x^4-5\right) \log \left(\frac{1}{2}
   \left(\sqrt{\frac{1}{x^2}-4}+\frac{1}{x}\right)\right)\cr 
   &+\left(
   4 x^2-1\right) \left(-344 x^8+416 x^6-201 x^4+30 x^2+6
   \left(32 x^8-128 x^6+114 x^4-41 x^2+5\right) \log
   (x)\right)\Bigg] \, , \\
g_{{LL}}(x) &=  \frac{-244 x^6+268 x^4-123 x^2+18}{x^4-4 x^6}+\frac{6 x
   \left(\frac{28 x^4-19 x^2+3}{x^6}-2\right)
   \left(\frac{1}{x^2}-2\right)^2 \log \left(\frac{1}{2}
   \left(\sqrt{\frac{1}{x^2}-4} + \frac{1}{x}\right)\right)}{\left
   (\frac{1}{x^2}-4\right)^{3/2}} \cr &+\frac{6 \left(-4 x^6+20 x^4-13
   x^2+3\right) \log (x)}{x^6} \, .   
\end{align}
}
We plot the functions $g_{TT, TL, LL}$ in Fig.\,\ref{fig:gfun}.
These functions are real and positive for any real values of $m_V$ and $m_\phi$, and we find that the gravitational contributions from the Higgs to $B(\Lambda)$ are negative. 

In the limit of $m_V/m_\phi \to 0$ and $\to \infty$, the above expressions reduce to
\begin{align}
    B^{TT}_{\text{grav}} =& \begin{dcases}
    -\frac{g_\Phi^2}{72 \pi ^2 m_\phi^2\Mpl^2} &
    (m_V/m_\phi \to 0)\,, \\
    -\frac{g_\Phi^2}{24 \pi  m_{V} m_\phi \Mpl^2}  &
    (m_V/m_\phi \to \infty)\,,
    \end{dcases}  \\
     B^{TL}_{\text{grav}} =& \begin{dcases}
    -\frac{g_\Phi^2}{144 \pi ^2 m_{V}^2 \Mpl^2}  &
    (m_V/m_\phi \to 0)\,, \\
    -\frac{g_\Phi^2}{24 \pi  m_{V} m_\phi \Mpl^2}  &
    (m_V/m_\phi \to \infty)\,,
    \end{dcases}  \\
     B^{LL}_{\text{grav}} =& \begin{dcases}
   -\frac{g_{\Phi}^2}{72 \pi ^2 m_{V}^2 \Mpl^2}  &
   (m_V/m_\phi \to 0)\,, \\
    -\frac{g_\Phi^2}{24 \pi  m_{V} m_\phi \Mpl^2}  & 
    (m_V/m_\phi \to \infty)\,.
    \end{dcases} 
\end{align}

\subsection{Fermion Contribution}
Next, let us introduce a dark $\mathrm{U}(1)$ charged Dirac fermion $\psi$ as a matter field, whose interaction is given by
\begin{align}
    \mathcal{L}_{\psi} = i \bar{\psi} \Slash{D} \psi - m_F \bar{\psi} \psi \ , \label{eq:fermion_int}
\end{align}
where $D_{\mu} = \partial_{\mu} - ig_{F} V_{\mu} $ is the covariant derivative.
Note that this interaction can destabilize the gauge boson, by allowing $V \to \psi \bar{\psi}$ decay if $m_V  > 2 m_F$ and decay into gravitons.

If the mass of the charged particles is large $(m_F\gg \Lambda)$, these charged particles are integrated out in the EFT and yield 
higher-derivative corrections to the EFT action. These effects to $B(\Lambda)$ can be estimated as $\mathcal{O}(g_F^4/m_F^4)$ and we can ignore them compared to contributions \eqref{BTTng} from the Higgs field. Hence, our interest is the charged particle within the regime of EFT ($m_F < \Lambda$).

\subsubsection{Calculation of $B(\Lambda)$}

\begin{figure}[t!]
	\centering
 	\subcaptionbox{Diagram for $B_{\text{non-grav}}$.\label{fig:nongra_f}}
	{\includegraphics{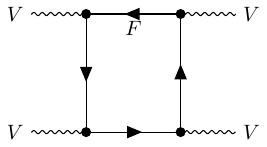}}
 \qquad
 	\subcaptionbox{Diagram for $B_{\text{grav}}$.\label{fig:gra_f}}
	{\includegraphics[trim = -20 0 0 0]{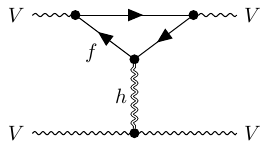}}
 \caption{The fermion loop contributions to $B$.}
 \label{fig:Higgs-loop}
\end{figure}	
We calculate the contributions of fermion-loop diagrams to $B(\Lambda)$ for the dark photon scattering amplitudes at one-loop order.

\subsubsection*{Non-gravitational contributions}
The diagram for a non-gravitational fermion loop is shown in Fig.\,\ref{fig:nongra_f}.
The one-loop contributions are given by
\begin{align}
    B_{\text{non-grav},F}^{TT}&= \frac{g_F^4 \left( 2 \log \frac{\Lambda^2}{m_F^2} + 1 \right)}{4\pi^2 \Lambda^4} +\cdots \, ,\\
    B_{\text{non-grav},F}^{TL}&=\frac{4 g_{F}^4 m_{V}^{2}}{3\pi^2\Lambda^6} +\cdots \, ,\\
    B_{\text{non-grav},F}^{LL}&=\frac{g_F^4 m_{V}^{4} \left( 4\log \frac{\Lambda^2}{m_{F}^2} - 7 \right)}{\pi^2\Lambda^8}    +\cdots
\end{align}
for $\Lambda \gg m_{V,F}$.
These contributions of $LL$ and $TL$ vanish in the limit $m_V/\Lambda \to 0$. 
Recall that $B_{\text{non-grav}}$ is determined by the high-energy limit of the forward limit amplitude thanks to the dispersion relation. 
The behaviors $B_{\text{non-grav},F}^{TL}, B_{\text{non-grav},F}^{LL} \to 0$ as $m_V/\Lambda \to 0$ can be understood by the decoupling of the longitudinal sector in the high-energy limit. The contributions of the charged particles to $B_{\text{non-grav}}^{TL}$ and $B_{\text{non-grav}}^{LL}$ can be ignored in the limit $\Lambda \gg m_V$. 

We thus focus on the $TT$ scattering.
The charged spin-1/2 loops give positive contributions to the non-gravitational diagram of the order $B_{\text{non-grav},F}^{TT} \sim g_F^4/\Lambda^4$, which is of the same order as the Higgs counterpart \eqref{BTTng}. Therefore, the inclusion of charged spin-1/2 particles does not drastically change the non-gravitational part. We reach a similar conclusion even in the presence of charged spin-0 particles~\cite{Alberte:2020bdz}. 

In contrast, as studied in~\cite{Aoki:2021ckh, Noumi:2022zht}, charged spin-1 loops with the mass $m_{\text{spin-1}}$ and the charge $g_{\text{spin-1}}$ lead to a different asymptotic behavior $B_{\text{non-grav}}^{TT} \sim g_{\text{spin-1}}^4/(m_{\text{spin-1}}^2 \Lambda^2)$ which dominates over other contributions in the limit $\Lambda \gg m_{\text{spin-1}}$. One may then wonder about contributions from higher-spin particles. The scaling in $\Lambda$ is determined by the high-energy behavior of the imaginary part of the amplitude in the forward limit, i.e.,~the total cross-section. Roughly, spin-0, 1/2 loops give $\mathrm{Im}\,\mathcal{M} \propto s^0$ while spin-1 loops yield $\mathrm{Im}\,\mathcal{M} \propto s/m_{\text{spin-1}}^2$ in high-energy limit. The faster growth in $s$ provides a larger contribution to $B_{\text{non-grav}}(\Lambda)$. However, the asymptotic growth of the cross-section in $s$ is bounded by the Froissart bound \cite{Froissart:1961ux} as $\mathrm{Im}\,\mathcal{M} < s \ln^2 s$ in gapped theories, implying that the spin-1 contribution almost saturates the bound. Higher-spin particles, if they are described by gapped theories like QCD, may not give drastically different contributions compared to  spin-1 particles. For instance, in the case of light-by-light scattering, the hadronic contribution can be estimated as $\mathrm{Im}\,\mathcal{M} \propto s^{1.08}$ by employing the vector meson dominance model~\cite{Donnachie:1992ny,Klusek:2009yi}.

In short, if a charged spin-1 or higher-spin particle appears at $m$, the non-gravitational contribution $B_{\text{non-grav}}^{TT}$ needs to be modified in $\Lambda > m$. By contrast, spin-0 or spin-1/2 particles give additional contributions but the qualitative behavior of $B_{\text{non-grav}}^{TT}$ is the same. 

\subsubsection*{Gravitational contributions}
\begin{figure}[t!]
    \centering
    \includegraphics[width=100mm]{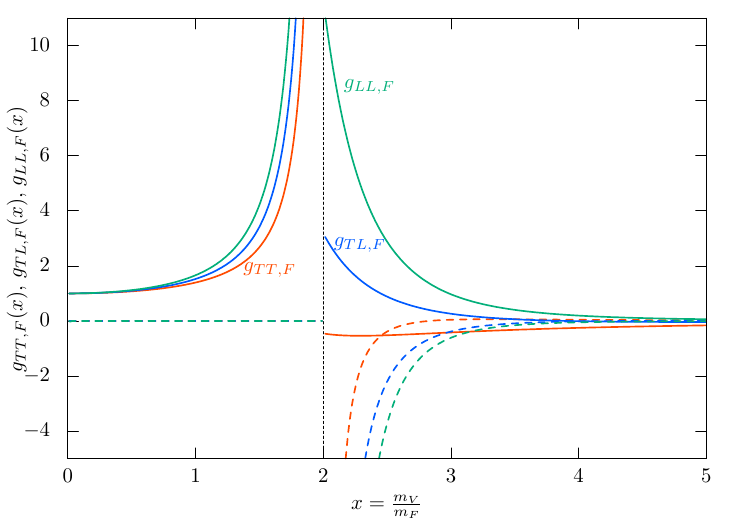}
    \caption{The functions $g_{{TT},F}(x)$ (red), $g_{{TL},F}(x)$ (green) and $g_{{LL},F}(x)$ (blue).
    The solid and dashed lines represent the real and imaginary parts, respectively.
    }
    \label{fig:g_plot}
\end{figure}

Next, we discuss the gravitational parts (Fig.\,\ref{fig:gra_f}), which are given by
\begin{align}
    B_{\mathrm{grav},F}^{TT} &= -\frac{11 g_{F}^2}{360\pi^2 \Mpl^2 m_F^2 } \,g_{{TT},F} \left( \frac{m_V}{m_F} \right),
    \\
    B_{\mathrm{grav},F}^{TL} &= -\frac{{11} g_{F}^2}{720\pi^2 \Mpl^2 m_F^2 }\, g_{{TL},F} \left( \frac{m_V}{m_F} \right),
    \\
    B_{\mathrm{grav},F}^{LL} &= -\frac{g_{F}^2 m_V^2}{420 \pi^2 \Mpl^2 m_F^4 } \, g_{{LL},F} \left( \frac{m_V}{m_F} \right),
\end{align}
where 
\begin{align}
     g_{{TT},F} (x) &= \frac{60 \left(x^2-8\right) \left(x^2-2\right) \log
   \left(\frac{1}{2} x \left(\sqrt{x^2-4}-x\right)+1\right)-20x
   \sqrt{x^2-4} \left(2 x^4-11 x^2+24\right)}{11 x^5
   \left(x^2-4\right)^{3/2}}
   \,,
   \\
     g_{{TL},F} (x) &=\frac{60 \left(x^4-22 x^2+56\right) \log \left(\frac{1}{2} x
   \left(\sqrt{x^2-4}-x\right)+1\right)-20 x \sqrt{x^2-4}
   \left(x^4-19 x^2+84\right)}{11 x^5 \left(x^2-4\right)^{3/2}}
   \,,
   \\
   g_{{LL},F} (x) & =\frac{70 \left(\left(120-36 x^2\right) \log \left(\frac{1}{2} x
   \left(\sqrt{x^2-4}-x\right)+1\right)+x \sqrt{x^2-4}
   \left(x^4+8 x^2-60\right)\right)}{3 x^7
   \left(x^2-4\right)^{3/2}} \, .
\end{align}
We plot the functions $g_{AB,F}(x)~(A,B={T}, {L})$ in Fig.~\ref{fig:g_plot}.

Let us first discuss the limit $m_{V}\to 0$, which yields
    \begin{align}
             B_{\text{grav},F}^{TT} &= -\frac{11g_F^2 }{360\pi^2 M_{\text{pl}}^2 m_F^2}
             \,, \\
             B_{\text{grav},F}^{TL} &= -\frac{11g_F^2 }{720\pi^2 M_{\text{pl}}^2 m_F^2}
             \,, \\
             B_{\text{grav},F}^{LL} &= -\frac{g_F^2 m_{V}^2}{420\pi^2 M_{\text{pl}}^2 m_F^4}
             \,.
    \end{align}
If the Higgs and the gauge boson are lighter than the charged particles $(m_{\phi}, m_V \ll m_F)$, we can neglect these contributions to the gravitational process~$B_{\text{grav}}$ in comparison to the contributions of the gauge boson sector. Note that the $LL$ part has an additional suppression $m_V^2/m_F^2$ because the longitudinal mode should decouple from the charged particle in the limit $m_V \to 0$. 
\red{However, since the longitudinal mode does not decouple from gravity, the $TL$ mode does not have such a suppression factor in the diagram where the longitudinal mode attaches to the graviton line.}

By contrast, a peculiar behavior appears in a heavy mass range of the gauge boson. The functions $g_{AB,F}$ are singular at $m_V=2m_F$ and become complex numbers in $m_V > 2 m_F$. 
The threshold $m_V=2m_F$ corresponds to the value at which the decay of the gauge boson to the charged particles starts to be kinematically allowed. 
The singularity at $m_V=2m_F$ is understood as an anomalous threshold. 
Let us consider the triangle diagram of Fig.~\ref{fig:gra_f}, which gives rise to the gravitational form factor of the gauge boson. The triangle diagram has a normal threshold at $t=4m_F^2$. In addition, the anomalous threshold comes up on the first sheet of the complex $t$-plane in $m_V > \sqrt{2}m_F$ and the position of the singularity is at
\begin{align}
  t=\frac{m_V^2}{m_F^2}(4m_F^2-m_V^2)
\,.
\end{align}
Therefore, the form factor at $t=0$ is singular when $m_V=2m_F$, generating the singularity of $B_{\text{grav}, F}^{AB}$. In Fig.~\ref{fig:g_plot}, we have plotted the functions in the unstable range $m_V > 2m_F$ as well. Recall that $B(\Lambda)$ has to be real according to the dispersion relation \eqref{B_Lambda} if all the mentioned properties are satisfied. 
This implies that at least one of the properties does not hold when the gauge boson decays. 
In fact, amplitudes exhibit peculiar behaviors if the mass of the external particle is extrapolated to the unstable region~\cite{Hannesdottir:2022bmo,Aoki:2022qbf}. The conventional positivity bounds need to be modified for unstable particles.

We expect that the subtleties associated with the decay are negligible if the particle is long-lived ($g_F \ll 1$). In other words, although the fermion contribution to $B_{\rm grav}(\Lambda)$ is quite subtle in 
the mass range $m_V > 2m_F$, this subtle contribution can be smaller than the contribution from the Higgs sector and could be simply negligible. In the following, we shall adopt this optimistic expectation when discussing models in which the gauge boson decays into other particles. The issue of unstable particles will be studied elsewhere.

\subsection{Constraint on Dark Gauge Boson}

\begin{table}[t]
  \centering
  \caption{Summary of the gravitational and non-gravitational contributions to $B$ for $m_V \ll m_{\phi, F} \ll \Lambda$ and $m_\phi \ll \sqrt{\Lambda m_V}$. }
  \label{tab:B}
  \begin{tabular}{|c|c|c|c|c|}
    \hline
   & \multicolumn{2}{|c|}{Higgs loop} & \multicolumn{2}{|c|}{Fermion  loop} \\
    \hline
   & $B_{\text{non-grav}}$ &  $B_{\text{grav}}$  & $B_{\text{non-grav}}$ &  $B_{\text{grav}}$  \\
    \hline
    $TT$& $\dfrac{g_\Phi^4}{4 \pi ^2 \Lambda ^4}$ & $-\dfrac{g_\Phi^2}{72 \pi ^2 \Mpl^2 m_\phi^2}$ & $\dfrac{g_F^4 \left( 2 \log \frac{\Lambda^2}{m_F^2} + 1 \right)}{4\pi^2 \Lambda^4}$ & $-\dfrac{11g_F^2 }{360\pi^2 \Mpl^2 m_F^2}$ \\
    \hline
    $TL$ & $\dfrac{g_\Phi^4}{2 \pi ^2 \Lambda ^4}$ & $-\dfrac{g_\Phi^2}{144 \pi ^2 \Mpl^2  m_{V}^2} $ & $\dfrac{4 g_{F}^4 m_{V}^{2}}{3\pi^2\Lambda^6}$ & $-\dfrac{11g_F^2 }{720\pi^2 \Mpl^2 m_F^2}$ \\
    \hline
    $LL$ & $\dfrac{g_\Phi^4}{\pi ^2 \Lambda ^2 m_{V}^2}$ & $-\dfrac{g_{\Phi}^2}{72 \pi ^2  \Mpl^2 m_{V}^2} $ & $\dfrac{g_F^4 m_{V}^{4} \left( 4\log \frac{\Lambda^2}{m_{F}^2} - 7 \right)}{\pi^2\Lambda^8}$ & $ -\dfrac{g_F^2 m_{V}^2}{420\pi^2 \Mpl^2 m_F^4}$ \\    
    \hline
  \end{tabular}
\end{table}

In Table \ref{tab:B}, we summarize our estimation of $B(\Lambda)$ for a light gauge boson.
\red{
We apply the gravitational positivity bound \eqref{g-positivity} to these amplitudes:
    \begin{align}\label{g-positivity_gb}
        B(\Lambda) = B_{\rm non-grav}(\Lambda) + B_{\rm grav}(\Lambda) \ge \frac{\mysigma}{\Mpl^2 M^2} \,.
    \end{align}
    
For simplicity of the presentation, in the following, we will discuss the constraints on the dark gauge boson parameters for $\mysigma=0$ with a fixed value of $\Lambda$ (Interpretation \ref{interpretation1} in Sec.~\ref{subsection:g_bounds}).
One should recall, however, that the actual bound is \eqref{g-positivity_gb}.
The bound becomes stronger for $\sigma=+1$ and weaker for $\sigma=-1$. 
We will later see more quantitatively how the $\sigma/(\Mpl^2 M^2)$ term changes the bound in Fig.~\ref{fig:M}.
}


\subsubsection*{Abelian Higgs without matter fields}

\begin{figure}[t!]
	\centering
 	\subcaptionbox{\label{fig:Higgs}}
	{\includegraphics[width=0.49\textwidth]{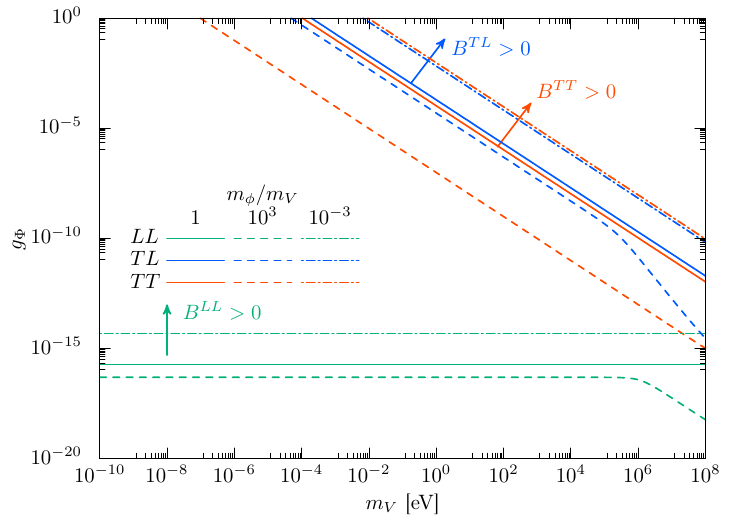}}
 	\subcaptionbox{\label{fig:mv_mh}}
	{\includegraphics[width=0.49\textwidth]{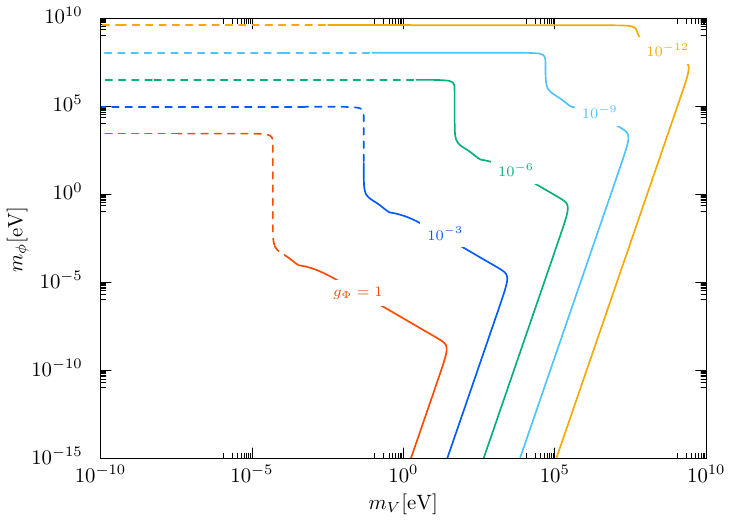}}
 \caption{The constraints on the dark gauge boson parameters in the Higgs mechanism.
  (a): Lower limit of $g_\Phi$ as a function of $m_V$. The positivity bound is satisfied in the region above the lines.
  (b): Contour of the lower bound of $g_\Phi$  on $m_V-m_\phi$ plane from the positivity bound. The dashed lines indicate the Higgs quartic coupling is large $\lambda > \pi$.
  We choose $\Lambda = 1$ TeV in both figures. 
 }
 \label{fig:constraints}
\end{figure}

Let us first look at the case where there is no matter field and only gauge and Higgs fields exist.
For a very light gauge boson $m_V \ll m_{\phi, F} \ll \Lambda$ and $m_\phi \ll \sqrt{\Lambda m_V}$, the following conditions can be obtained from each helicity
\begin{align}
    B^{TT}_{\text{non-grav}} + B^{TT}_{\text{grav}} \ge 0 \quad \longrightarrow &  \quad  m_{\phi} \ge \frac{ \Lambda^2 }{3\sqrt{2} g_\Phi  \Mpl} \, ,\\
    B^{TL}_{\text{non-grav}} + B^{TL}_{\text{grav}} \ge 0 \quad \longrightarrow &  \quad m_V \ge \frac{\Lambda^2 }{6 \sqrt{2} g_\Phi  \Mpl}  \, \label{eq:TLcons},\\    
    B^{LL}_{\text{non-grav}} + B^{LL}_{\text{grav}} \ge 0 \quad \longrightarrow &  \quad g_\Phi \ge \frac{\Lambda}{6 \sqrt{2} \Mpl}  \,.        \label{eq:LLcons}
 \end{align}
In Fig.\,\ref{fig:constraints}, we show the positivity conditions. 
Note that the higher terms of $\Lambda$ for the non-gravitational contributions can be dominant if $m_\phi \gtrsim  \sqrt{m_V \Lambda}$ or $m_\phi \lesssim  m_V^2/ \Lambda$.

\subsubsection*{St\"{u}ckelberg gauge boson with matter field}

\begin{figure}[t!]
	\centering

	\includegraphics[width=0.49\textwidth]{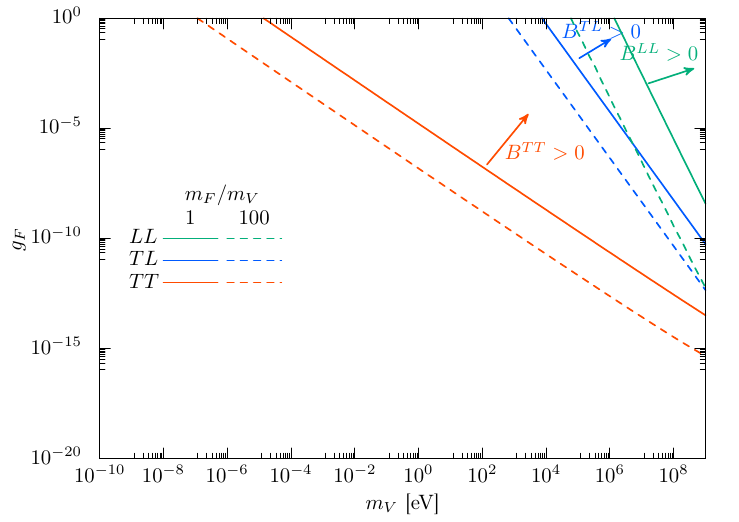}
 \caption{The constraints on the dark gauge boson parameters for the St\"{u}ckelberg  $\mathrm{U}(1)$ gauge theory with a fermion field.
The positivity bound is satisfied in the region above the lines.
  We choose $\Lambda = 1$ TeV.
 }
 \label{fig:st_constraints}
\end{figure}

Next, let us discuss the case where the Higgs field $\phi$ is absent in the EFT, as in the case of the St\"{u}ckelberg mechanism. We find
%
%
\begin{align}
    B^{TT}_{\text{non-grav}} + B^{TT}_{\text{grav}} \ge 0 \quad \longrightarrow &  \quad  g_F \gtrsim 0.2 \frac{\Lambda^2}{m_F \Mpl \sqrt{\log(\Lambda m_F^{-1})}} \, ,\\
    B^{TL}_{\text{non-grav}} + B^{TL}_{\text{grav}} \ge 0 \quad \longrightarrow &  \quad m_V  \gtrsim 0.1 \frac{\Lambda^3 }{ g_F m_F  \Mpl}\, ,\\    
    B^{LL}_{\text{non-grav}} + B^{LL}_{\text{grav}} \ge 0 \quad \longrightarrow &  \quad m_V \gtrsim 0.02  \frac{\Lambda^4}{g_F m_F^2 \Mpl \sqrt{\log(\Lambda m_F^{-1})}} \, .         
 \end{align}
In Fig.\,\ref{fig:st_constraints}, we show the positivity conditions on $m_V - g_F$ plane.
Note that for $m_V > 2 m_F$, delicate arguments are required and the positivity conditions used previously cannot be applied as they are.

More generally, it is possible to incorporate both Higgs and fermions. 
The helicity configurations involving longitudinal modes provide stronger constraints.
As discussed earlier, the Higgs contribution is dominant when $g_\Phi \sim g_F$ and $m_V \ll m_{\phi,F}$, in which case the positivity conditions will be similar to those of the Higgs scenarios.

\subsubsection*{Comparison with other swampland constraints for St\"{u}ckelberg gauge bosons}

\begin{figure}[htbp]
	\centering
 	\subcaptionbox{Light fermion\label{fig:light_fermion}}
	{\includegraphics[width=0.49\textwidth]{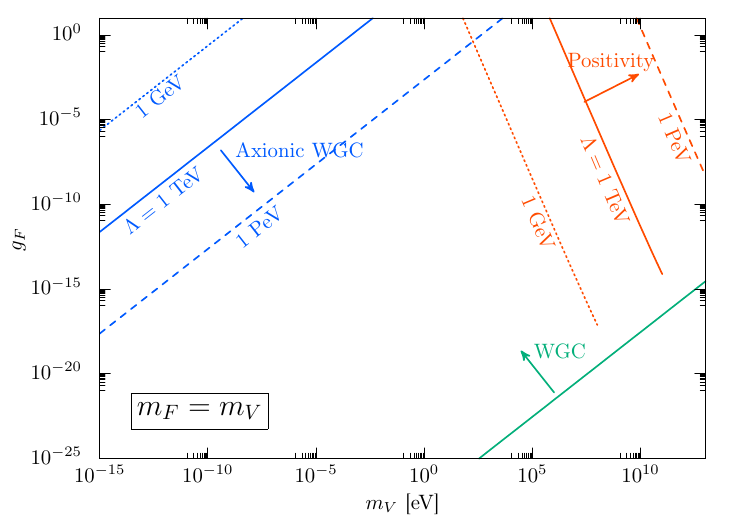}}
 	\subcaptionbox{Heavy fermion\label{fig:heavy_fermion}}
	{\includegraphics[width=0.49\textwidth]{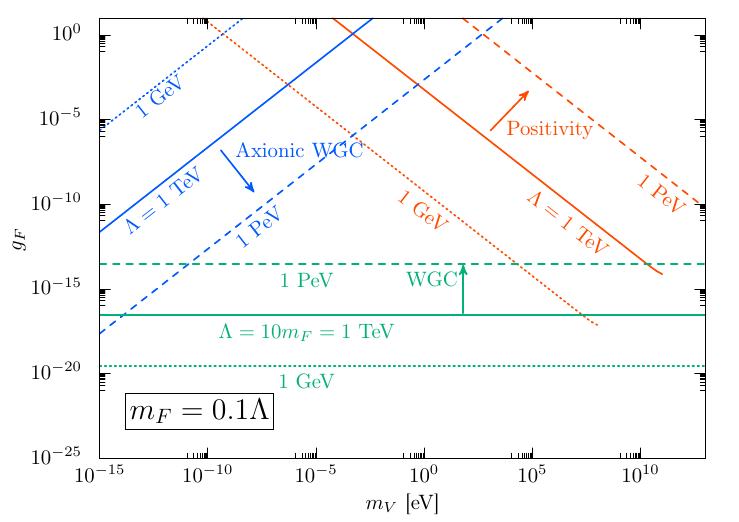}}
\caption{Comparison between the swampland constraints for St\"{u}ckelberg gauge bosons, the gravitational positivity bound (in red) and other swampland bounds \eqref{bound_Reece} (in blue) and the weak gravity conjectures (in green).
We take the cutoff scale $\Lambda$ is 1 GeV (dotted lines), 1 TeV (solid lines) and 1 PeV (dashed lines).
(a): Light fermion case $m_F = m_V$ and (b): Heavy fermion case $m_F = 0.1 \Lambda$.
In this plot, the lines are chopped in the parameter space where  $m_V >0.1 \Lambda$. 
}
\label{fig:comparison}
\end{figure}

Let us compare our swampland constraints with another swampland constraint on the gauge boson mass \cite{Reece:2018zvv} (see also \cite{Hebecker:2017uix}) motivated by the swampland distance conjecture \cite{Ooguri:2006in}.
This bound states that the UV cutoff $\Lambda_{\rm UV}$ of an EFT with a gauge boson should obey 
\begin{align} \label{bound_Reece}
\Lambda_{\rm UV} \lesssim \textrm{min}( (m_V \Mpl/g)^{1/2}, g^{1/3} \Mpl) \, ,
\end{align}
where (as before) $m_V$ is the gauge boson mass and $g$ is the gauge coupling constant.
In Fig.~\ref{fig:comparison} we compare our bounds and the bound \eqref{bound_Reece}
for sample values of parameters.

One can see in Fig.~\ref{fig:comparison} that overall our bound is stronger than the bound \eqref{bound_Reece}.
Let us quickly point out, however, that the two bounds are derived under a different set of assumptions/arguments,
and one needs to be careful in any meaningful comparison. 

The gravitational positivity bound is
derived by general properties of Reggeized scattering amplitudes of a weakly-coupled UV theory.
The assumptions are believed to be standard, but
the bound has some ambiguities due to the unknown UV mass scale/sign $(M, \mysigma)$ in \eqref{M_def}.
By contrast, the bound \eqref{bound_Reece} is derived from a set of swampland conjectures.
The $(m_V \Mpl/g)^{1/2}$ bound in \eqref{bound_Reece} is derived by noting that the St\"{u}ckelberg mass is obtained when the gauge boson eats a fundamental axion,
for which we can apply an axionic version \cite{Hebecker:2017uix,Reece:2018zvv} of the weak gravity conjecture \cite{Arkani-Hamed:2006emk};
the $g^{1/3} \Mpl$ bound in \eqref{bound_Reece} is derived from a combination of the (sub)lattice/tower weak gravity conjecture \cite{Heidenreich:2016aqi,Heidenreich:2017sim} (see also \cite{Montero:2016tif,Andriolo:2018lvp}) and the species bound \cite{Arkani-Hamed:2005zuc,Dvali:2007hz,Dvali:2007wp}. 

Let us also note that our bound applies to gauge boson masses
both in the Higgs mechanism and the St\"{u}ckelberg mechanism,
while the bound \eqref{bound_Reece} applies only to those in the St\"{u}ckelberg mechanism.
Our bound is more general in this respect.

In Fig.~\ref{fig:comparison} we also plotted the bound from the weak gravity conjecture \cite{Arkani-Hamed:2006emk}, which requires an existence of a particle with charge $g$ and mass $m$ such that 
$\sqrt{2} g \ge  m/M_{\rm Pl}$.

\section{Toward Constraints on Realistic Phenomenological Models}\label{sec:toward}

The dark $\mathrm{U}(1)$ gauge sectors discussed so far have nothing to do with our universe and are entirely theoretical constructs.
However, any new physics  describing our universe must necessarily coexist with the Standard Model.
Since some important subtleties arise at this point, we will first briefly discuss them and then discuss the specific model (B$-$L gauge model) as a possible application of the gravitational positivity.

If we apply the discussion of gravitational positivity bounds to, for example, the scattering of photons and dark gauge bosons, the dark sector necessarily has interactions beyond those described by the Standard Model and gravity~\cite{Noumi:2022zht}.
In general, these interactions allow the dark gauge boson to decay into photons and SM fermions and also decay into gravitons.

\begin{figure}[t]
    \centering
    \includegraphics[width=120mm]{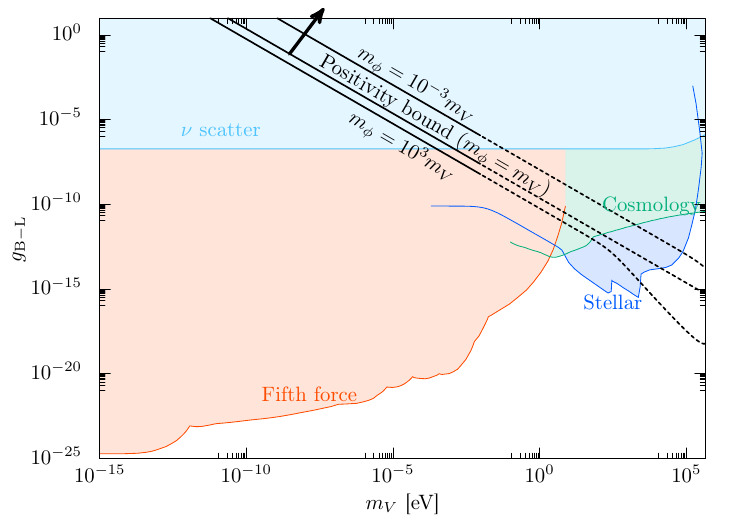}
    \caption{Implication to the $\mathrm{U}(1)_{\mathrm{B-L}}$ gauge boson from the gravitational positivity bound {\color{black} with $\sigma=0$.}
    Here we adopt $\Lambda = 1$ GeV.
    The positivity constraint excludes the parameter region below the black lines.
    The dotted lines indicate that the gauge boson mass is greater than neutrino masses and our analysis cannot be directly applied.
    We also show several experimental constraints from the fifth force searches \cite{Hoskins:1985tn,Smith:1999cr,Kapner:2006si, Wagner:2012ui,Lee:2020zjt,Yang:2012zzb,Tan:2020vpf,Chen:2014oda,MICROSCOPE:2022doy}, stellar cooling \cite{Hardy:2016kme}, cosmological constraints on the neutrino properties \cite{Escudero:2019gvw,Ibe:2020dly}, and XENONnT constraint on the $e-\nu$ scattering \cite{XENON:2022ltv,A:2022acy}. 
    We assume that right-handed neutrinos are heavy and not relevant to the constraints.
    }
    \label{fig:BLgauge}
\end{figure}

In this case, the dark gauge boson has a finite decay width, and hence care must be taken when applying the positivity constraint.
Despite those caveats, the decay rate of light and weakly interacting gauge bosons is extremely small, and we expect no major practical problems.

As an example, we consider the $\mathrm{U}(1)_{\mathrm{B-L}}$ extension of the Standard Model.
We assume the gauge charge of the B$-$L breaking Higgs is the same as that of the SM leptons.
In Fig.\,\ref{fig:BLgauge}, we show the positivity bound \eqref{g-positivity_gb} on the $\mathrm{U}(1)_{\mathrm{B-L}}$ model and current experimental constraints.
Here we take \red{$\sigma=0$ and} $\Lambda = 1$ GeV.
In this plot, we assume that the B$-$L Higgs contributions are dominated over the SM fermions and neglect the SM contributions.
The dashed lines in the figure show the parameter region where the dark gauge boson can decay into neutrinos.
The stringent bound from the positivity constraint comes from TL scattering mode.
Note that the constraint is more severe for the higher cutoff, as seen in the TL constraint \eqref{eq:TLcons}.
The gravitational positivity bound requires the larger gauge coupling for smaller masses and has strong tensions with the experimental searches.

\begin{figure}[t]
    \centering
    \includegraphics[width=120mm]{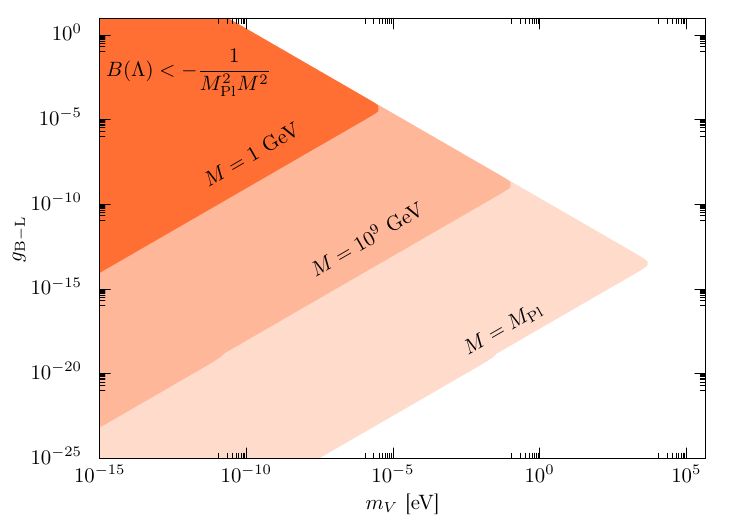}
    \caption{\red{ 
    Positivity constraints for various values of $M$ with $\sigma=-1$. The shaded regions do not satisfy $B(\Lambda) > \sigma/(\Mpl^2 M^2)$. Here, we set $m_V = m_\phi$ and $\Lambda = 1$ GeV.}
    }
    \label{fig:M}
\end{figure}

\section{Conclusion and Discussion}\label{sec:conclusion}

In this paper, we discussed the practical procedure for deriving 
gravitational positivity bounds. We illustrated the procedure for dark gauge bosons,
and 
\textcolor{black}{discussed implications of the bounds.}

There are two complementary comments regarding our results.
First and foremost, our bound is very strong---in Fig.\,\ref{fig:BLgauge} most of the parameter regions 
unexplored by experiments are already excluded by the positivity bound.
This is a clear demonstration of the power of the gravitational positivity bound, and 
we believe that the gravitational positivity bound can give rather strong constraints for many EFTs in BSM physics.
It is a fascinating question to explore this point further.

\red{Despite this optimism, 
we should also be well aware of the fine prints and limitations of our results, and 
we should be careful in interpreting the results. 
Our bound is derived from several assumptions of UV physics as spelled out in Sec.~\ref{sec:recipe},
which can be violated in some UV scenarios.
In particular, we have assumed $\sigma=0$ in the gravitational positivity bound $B(\Lambda) > \sigma/(\Mpl^2 M^2)$. 
As we have remarked before, the bound becomes stronger for $\sigma=+1$ and weaker for $\sigma=-1$. For illustrative purposes, we show the bound $B(\Lambda) > \sigma/(\Mpl^2 M^2)$ for various values of $M$ with $\sigma=-1$ in Fig.~\ref{fig:M}. While the bound $B(\Lambda)>0$ is accurate when the scale $M$ is sufficiently large, the constraint region shrinks as $M$ decreases.\footnote{In Fig.~\ref{fig:M}, the small $g_{\mathrm{B-L}}$ is allowed because we have fixed the value of $M$ independently of the gauge coupling. 
It would be possible that $M$ varies with the IR data $(g_{\mathrm{B-L}}, m_V)$. 
It should be also remarked that, in the small-coupling limit, we might need to take into account higher-dimensional operators that have been neglected in the evaluation of $B(\Lambda)$. In these cases, the region might have a more complicated shape.} 
Further theoretical studies are required to carve out the parameter space of the UV data $(M,\sigma)$.
}
Conversely, if the violation of the inequality $B(\Lambda)>0$ is experimentally verified, it can be regarded as an IR constraint on quantum gravity (Interpretation \ref{interpretation3}). 
Moreover, there are other subtleties when we wish to apply the bounds to realistic models, as discussed in Sec.~\ref{sec:toward}.
We expect in general that the dark photon interacts with the Standard Model via kinetic mixings with the Standard-Model gauge bosons.
This means that there are large non-gravitational amplitudes from QCD that can have a significant impact on the dark photon constraint.
Another issue is the stability of the dark photon.
The dark photon is expected to decay into Standard Model fermions and photons once it interacts with the SM sector.
In this case, the dark photon has a finite decay width, and care must be taken when applying the positivity constraint.
It requires careful theoretical analysis to fully address these details, which we hope will be discussed in future works.

\section*{Acknowledgements}

This work was partly performed during our mutual visits to 
Kobe University, Kyoto University (Yukawa Institute), University of Tokyo (Kavli IPMU), and Yamaguchi University.
We would like to thank the institutes for their hospitality.
M.Y.\ is grateful to the Kavli Institute for Theoretical Physics in Santa Barbara for hospitality during the Integrable22 workshop.

The Feynman diagrams in this paper were drawn with the help of \texttt{TikZ-FeynHand}\, \cite{Ellis:2016jkw,Dohse:2018vqo}.

The work of K. A. was supported in part by Grant-in-Aid from the Scientific Research Fund of the Japan Society for the Promotion of Science, No. 20K14468 and No. 24K17046.
TN was supported in part by JSPS KAKENHI Grant No. 20H01902 and No. 22H01220, and MEXT KAKENHI Grant
No. 21H05184 and No. 23H04007. 
This work of RS is supported in part by JSPS Grant-in-Aid for Scientific
Research 20H05860 and No.\ 23H01171.
S.~Sato is supported in part by JST SPRING Grant No. JPMJFS2126. 
The work of S.~Shirai is supported by Grant-in-Aid for Scientific Research from the Ministry of Education, Culture, Sports, Science, and Technology (MEXT), Japan, 18K13535, 20H01895, 20H05860 and 21H00067, and by World Premier International Research Center Initiative (WPI), MEXT, Japan. 
The work of JT was supported by IBS under the project code, IBS-R018-D1.
The work of MY is supported in part by the JSPS Grant-in-Aid for Scientific Research (No.\ 19H00689, 19K03820, 20H05860, 23H01168) and by JST, Japan (PRESTO Grant No.\ JPMJPR225A and Moonshot R\&D Grant No.\ JPMJMS2061). 

\appendix
\section{\texorpdfstring{Gravitational Contributions Beyond Graviton $t$-channel}{Gravitational Contributions Beyond Graviton t-channel}} \label{sec:non_grav}
\begin{figure}[h!]
	\centering
	{\includegraphics{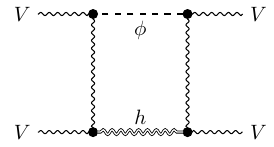}}
 \caption{An example of a gravitational contribution beyond $t$-channel exchange. \label{fig:others}}
\end{figure}	

The gravitational contribution $B_{\text{grav}}(\Lambda)$ to $B(\Lambda)$ can be divided into
 $B_{\text{grav,$t$-channel}}(\Lambda)$ from the graviton $t$-channel exchange
 and the remaining part  $B_{\text{grav,others}}(\Lambda)$ (see Fig.\,\ref{fig:others}).
In this Appendix we show that we can neglect $B_{\text{grav,others}}(\Lambda)$ in comparison with $B_{\text{grav,$t$-channel}}(\Lambda)$.
  
 Calculation of $B_{\text{grav,others}}(\Lambda)$ is challenging due to the presence of IR-divergent diagrams in $\widetilde{\mathcal{M}}_{\text{grav,others}}(s,t)$. 
 While it is more desirable to define IR-finite amplitudes using the dressed state formalism, it is beyond the scope of this work and will be pursued in future research. 
 To regularize this IR divergence, instead, 
 we introduce an IR regulator $\mu_\mathrm{IR}$ by deforming the graviton propagator as
     \begin{align}
        \frac{i P_{\mu \nu \rho \sigma}}{q^2+i \epsilon} \rightarrow \frac{i P_{\mu \nu \rho \sigma}}{q^2-\mu_{\mathrm{IR}}^2+i \epsilon}\ ,
    \end{align}
where $P_{\mu \nu \rho \sigma}$ is defined by
$P_{\mu \nu \rho \sigma} =   \left( \eta_{\mu \rho} \eta_{\nu \sigma}+\eta_{\mu \sigma} \eta_{\nu \rho}- \eta_{\mu \nu} \eta_{\rho \sigma}\right) /2 $
in the harmonic gauge.   
Using the deformed graviton propagator, we can confirm that $|\widetilde{\mathcal{M}}_{\text{grav,others}}(s,t)|$ is bounded by $\mathcal{O}(s^6)$ as $|s|\to \infty$. 
Along with the analyticity of $\widetilde{\mathcal{M}}_{\text{grav,others}}$, 
this makes it possible to derive the relation
    \begin{align}
        B_{\text{grav, others}}(\Lambda) &=\frac{4}{\pi}\int_{\Lambda^2}^{\infty} ds\, \frac{{\rm Im}\, \widetilde{\mathcal{M}}_{\text{grav, others}}(s,t=0)}{\left(s-2m_V^2\right)^7}
        \ .
        \label{others0}
    \end{align}
Moreover, since $\textrm{Im}\, \widetilde{\mathcal{M}}_{\text{grav,others}}(s,t)$
is also bounded by $\mathcal{O}(s^6)$,  we obtain
    \begin{align}
        B_{\text{grav, others}}(\Lambda) &= \mathcal{O}\left( \frac{1}{M_{\text{Pl}}^2 \Lambda^2} \right)
        \ .
        \label{others1}
    \end{align}
This is much smaller compared to the contribution from the graviton $t$-channel exchange diagrams and can be safely neglected. 

\bibliographystyle{JHEP}
\bibliography{bibtex}

\providecommand{\href}[2]{#2}\begingroup\raggedright\begin{thebibliography}{10}

\bibitem{tHooft:1979rat}
G.~'t~Hooft, \emph{{Naturalness, chiral symmetry, and spontaneous chiral
  symmetry breaking}},
  \href{https://doi.org/10.1007/978-1-4684-7571-5_9}{\emph{NATO Sci. Ser. B}
  {\bfseries 59} (1980) 135}.

\bibitem{Vafa:2005ui}
C.~Vafa, \emph{{The String landscape and the swampland}},
  \href{https://arxiv.org/abs/hep-th/0509212}{{\ttfamily hep-th/0509212}}.

\bibitem{Ooguri:2006in}
H.~Ooguri and C.~Vafa, \emph{{On the Geometry of the String Landscape and the
  Swampland}},
  \href{https://doi.org/10.1016/j.nuclphysb.2006.10.033}{\emph{Nucl. Phys. B}
  {\bfseries 766} (2007) 21}
  [\href{https://arxiv.org/abs/hep-th/0605264}{{\ttfamily hep-th/0605264}}].

\bibitem{Pham:1985cr}
T.N.~Pham and T.N.~Truong, \emph{{Evaluation of the Derivative Quartic Terms of
  the Meson Chiral Lagrangian From Forward Dispersion Relation}},
  \href{https://doi.org/10.1103/PhysRevD.31.3027}{\emph{Phys. Rev. D}
  {\bfseries 31} (1985) 3027}.

\bibitem{Pennington:1994kc}
M.R.~Pennington and J.~Portoles, \emph{{The Chiral Lagrangian parameters, l1,
  l2, are determined by the rho resonance}},
  \href{https://doi.org/10.1016/0370-2693(94)01551-M}{\emph{Phys. Lett. B}
  {\bfseries 344} (1995) 399}
  [\href{https://arxiv.org/abs/hep-ph/9409426}{{\ttfamily hep-ph/9409426}}].

\bibitem{Ananthanarayan:1994hf}
B.~Ananthanarayan, D.~Toublan and G.~Wanders, \emph{{Consistency of the chiral
  pion pion scattering amplitudes with axiomatic constraints}},
  \href{https://doi.org/10.1103/PhysRevD.51.1093}{\emph{Phys. Rev. D}
  {\bfseries 51} (1995) 1093}
  [\href{https://arxiv.org/abs/hep-ph/9410302}{{\ttfamily hep-ph/9410302}}].

\bibitem{Comellas:1995hq}
J.~Comellas, J.I.~Latorre and J.~Taron, \emph{{Constraints on chiral
  perturbation theory parameters from QCD inequalities}},
  \href{https://doi.org/10.1016/0370-2693(95)01110-C}{\emph{Phys. Lett. B}
  {\bfseries 360} (1995) 109}
  [\href{https://arxiv.org/abs/hep-ph/9507258}{{\ttfamily hep-ph/9507258}}].

\bibitem{Adams:2006sv}
A.~Adams, N.~Arkani-Hamed, S.~Dubovsky, A.~Nicolis and R.~Rattazzi,
  \emph{{Causality, analyticity and an IR obstruction to UV completion}},
  \href{https://doi.org/10.1088/1126-6708/2006/10/014}{\emph{JHEP} {\bfseries
  10} (2006) 014} [\href{https://arxiv.org/abs/hep-th/0602178}{{\ttfamily
  hep-th/0602178}}].

\bibitem{deRham:2022hpx}
C.~de~Rham, S.~Kundu, M.~Reece, A.J.~Tolley and S.-Y.~Zhou, \emph{{Snowmass
  White Paper: UV Constraints on IR Physics}},  in \emph{{Snowmass 2021}}, 3,
  2022 [\href{https://arxiv.org/abs/2203.06805}{{\ttfamily 2203.06805}}].

\bibitem{Hamada:2018dde}
Y.~Hamada, T.~Noumi and G.~Shiu, \emph{{Weak Gravity Conjecture from Unitarity
  and Causality}},
  \href{https://doi.org/10.1103/PhysRevLett.123.051601}{\emph{Phys. Rev. Lett.}
  {\bfseries 123} (2019) 051601}
  [\href{https://arxiv.org/abs/1810.03637}{{\ttfamily 1810.03637}}].

\bibitem{Bellazzini:2019xts}
B.~Bellazzini, M.~Lewandowski and J.~Serra, \emph{{Positivity of Amplitudes,
  Weak Gravity Conjecture, and Modified Gravity}},
  \href{https://doi.org/10.1103/PhysRevLett.123.251103}{\emph{Phys. Rev. Lett.}
  {\bfseries 123} (2019) 251103}
  [\href{https://arxiv.org/abs/1902.03250}{{\ttfamily 1902.03250}}].

\bibitem{Loges:2020trf}
G.J.~Loges, T.~Noumi and G.~Shiu, \emph{{Duality and Supersymmetry Constraints
  on the Weak Gravity Conjecture}},
  \href{https://doi.org/10.1007/JHEP11(2020)008}{\emph{JHEP} {\bfseries 11}
  (2020) 008} [\href{https://arxiv.org/abs/2006.06696}{{\ttfamily
  2006.06696}}].

\bibitem{Alberte:2020jsk}
L.~Alberte, C.~de~Rham, S.~Jaitly and A.J.~Tolley, \emph{{Positivity Bounds and
  the Massless Spin-2 Pole}},
  \href{https://doi.org/10.1103/PhysRevD.102.125023}{\emph{Phys. Rev. D}
  {\bfseries 102} (2020) 125023}
  [\href{https://arxiv.org/abs/2007.12667}{{\ttfamily 2007.12667}}].

\bibitem{Tokuda:2020mlf}
J.~Tokuda, K.~Aoki and S.~Hirano, \emph{{Gravitational positivity bounds}},
  \href{https://doi.org/10.1007/JHEP11(2020)054}{\emph{JHEP} {\bfseries 11}
  (2020) 054} [\href{https://arxiv.org/abs/2007.15009}{{\ttfamily
  2007.15009}}].

\bibitem{Herrero-Valea:2020wxz}
M.~Herrero-Valea, R.~Santos-Garcia and A.~Tokareva, \emph{{Massless positivity
  in graviton exchange}},
  \href{https://doi.org/10.1103/PhysRevD.104.085022}{\emph{Phys. Rev. D}
  {\bfseries 104} (2021) 085022}
  [\href{https://arxiv.org/abs/2011.11652}{{\ttfamily 2011.11652}}].

\bibitem{Alberte:2020bdz}
L.~Alberte, C.~de~Rham, S.~Jaitly and A.J.~Tolley, \emph{{QED positivity
  bounds}}, \href{https://doi.org/10.1103/PhysRevD.103.125020}{\emph{Phys. Rev.
  D} {\bfseries 103} (2021) 125020}
  [\href{https://arxiv.org/abs/2012.05798}{{\ttfamily 2012.05798}}].

\bibitem{Alberte:2021dnj}
L.~Alberte, C.~de~Rham, S.~Jaitly and A.J.~Tolley, \emph{{Reverse
  Bootstrapping: IR Lessons for UV Physics}},
  \href{https://doi.org/10.1103/PhysRevLett.128.051602}{\emph{Phys. Rev. Lett.}
  {\bfseries 128} (2022) 051602}
  [\href{https://arxiv.org/abs/2111.09226}{{\ttfamily 2111.09226}}].

\bibitem{Caron-Huot:2021rmr}
S.~Caron-Huot, D.~Mazac, L.~Rastelli and D.~Simmons-Duffin, \emph{{Sharp
  boundaries for the swampland}},
  \href{https://doi.org/10.1007/JHEP07(2021)110}{\emph{JHEP} {\bfseries 07}
  (2021) 110} [\href{https://arxiv.org/abs/2102.08951}{{\ttfamily
  2102.08951}}].

\bibitem{Caron-Huot:2022ugt}
S.~Caron-Huot, Y.-Z.~Li, J.~Parra-Martinez and D.~Simmons-Duffin,
  \emph{{Causality constraints on corrections to Einstein gravity}},
  \href{https://doi.org/10.1007/JHEP05(2023)122}{\emph{JHEP} {\bfseries 05}
  (2023) 122} [\href{https://arxiv.org/abs/2201.06602}{{\ttfamily
  2201.06602}}].

\bibitem{Herrero-Valea:2022lfd}
M.~Herrero-Valea, A.S.~Koshelev and A.~Tokareva, \emph{{UV graviton scattering
  and positivity bounds from IR dispersion relations}},
  \href{https://doi.org/10.1103/PhysRevD.106.105002}{\emph{Phys. Rev. D}
  {\bfseries 106} (2022) 105002}
  [\href{https://arxiv.org/abs/2205.13332}{{\ttfamily 2205.13332}}].

\bibitem{deRham:2022gfe}
C.~de~Rham, S.~Jaitly and A.J.~Tolley, \emph{{Constraints on Regge behavior
  from IR physics}},
  \href{https://doi.org/10.1103/PhysRevD.108.046011}{\emph{Phys. Rev. D}
  {\bfseries 108} (2023) 046011}
  [\href{https://arxiv.org/abs/2212.04975}{{\ttfamily 2212.04975}}].

\bibitem{Noumi:2022wwf}
T.~Noumi and J.~Tokuda, \emph{{Finite energy sum rules for gravitational Regge
  amplitudes}}, \href{https://doi.org/10.1007/JHEP06(2023)032}{\emph{JHEP}
  {\bfseries 06} (2023) 032}
  [\href{https://arxiv.org/abs/2212.08001}{{\ttfamily 2212.08001}}].

\bibitem{Hamada:2023cyt}
Y.~Hamada, R.~Kuramochi, G.J.~Loges and S.~Nakajima, \emph{{On (scalar QED)
  gravitational positivity bounds}},
  \href{https://doi.org/10.1007/JHEP05(2023)076}{\emph{JHEP} {\bfseries 05}
  (2023) 076} [\href{https://arxiv.org/abs/2301.01999}{{\ttfamily
  2301.01999}}].

\bibitem{Cheung:2014ega}
C.~Cheung and G.N.~Remmen, \emph{{Infrared Consistency and the Weak Gravity
  Conjecture}}, \href{https://doi.org/10.1007/JHEP12(2014)087}{\emph{JHEP}
  {\bfseries 12} (2014) 087} [\href{https://arxiv.org/abs/1407.7865}{{\ttfamily
  1407.7865}}].

\bibitem{Andriolo:2018lvp}
S.~Andriolo, D.~Junghans, T.~Noumi and G.~Shiu, \emph{{A Tower Weak Gravity
  Conjecture from Infrared Consistency}},
  \href{https://doi.org/10.1002/prop.201800020}{\emph{Fortsch. Phys.}
  {\bfseries 66} (2018) 1800020}
  [\href{https://arxiv.org/abs/1802.04287}{{\ttfamily 1802.04287}}].

\bibitem{Chen:2019qvr}
W.-M.~Chen, Y.-T.~Huang, T.~Noumi and C.~Wen, \emph{{Unitarity bounds on
  charged/neutral state mass ratios}},
  \href{https://doi.org/10.1103/PhysRevD.100.025016}{\emph{Phys. Rev. D}
  {\bfseries 100} (2019) 025016}
  [\href{https://arxiv.org/abs/1901.11480}{{\ttfamily 1901.11480}}].

\bibitem{Aoki:2021ckh}
K.~Aoki, T.Q.~Loc, T.~Noumi and J.~Tokuda, \emph{{Is the Standard Model in the
  Swampland? Consistency Requirements from Gravitational Scattering}},
  \href{https://doi.org/10.1103/PhysRevLett.127.091602}{\emph{Phys. Rev. Lett.}
  {\bfseries 127} (2021) 091602}
  [\href{https://arxiv.org/abs/2104.09682}{{\ttfamily 2104.09682}}].

\bibitem{Noumi:2021uuv}
T.~Noumi and J.~Tokuda, \emph{{Gravitational positivity bounds on scalar
  potentials}}, \href{https://doi.org/10.1103/PhysRevD.104.066022}{\emph{Phys.
  Rev. D} {\bfseries 104} (2021) 066022}
  [\href{https://arxiv.org/abs/2105.01436}{{\ttfamily 2105.01436}}].

\bibitem{Noumi:2022zht}
T.~Noumi, S.~Sato and J.~Tokuda, \emph{{Phenomenological motivation for
  gravitational positivity bounds: A case study of dark sector physics}},
  \href{https://doi.org/10.1103/PhysRevD.108.056013}{\emph{Phys. Rev. D}
  {\bfseries 108} (2023) 056013}
  [\href{https://arxiv.org/abs/2205.12835}{{\ttfamily 2205.12835}}].

\bibitem{Froissart:1961ux}
M.~Froissart, \emph{{Asymptotic behavior and subtractions in the Mandelstam
  representation}}, \href{https://doi.org/10.1103/PhysRev.123.1053}{\emph{Phys.
  Rev.} {\bfseries 123} (1961) 1053}.

\bibitem{Martin:1962rt}
A.~Martin, \emph{{Unitarity and high-energy behavior of scattering
  amplitudes}}, \href{https://doi.org/10.1103/PhysRev.129.1432}{\emph{Phys.
  Rev.} {\bfseries 129} (1963) 1432}.

\bibitem{Bellazzini:2020cot}
B.~Bellazzini, J.~Elias~Mir\'o, R.~Rattazzi, M.~Riembau and F.~Riva,
  \emph{{Positive moments for scattering amplitudes}},
  \href{https://doi.org/10.1103/PhysRevD.104.036006}{\emph{Phys. Rev. D}
  {\bfseries 104} (2021) 036006}
  [\href{https://arxiv.org/abs/2011.00037}{{\ttfamily 2011.00037}}].

\bibitem{Bellazzini:2021oaj}
B.~Bellazzini, M.~Riembau and F.~Riva, \emph{{IR side of positivity bounds}},
  \href{https://doi.org/10.1103/PhysRevD.106.105008}{\emph{Phys. Rev. D}
  {\bfseries 106} (2022) 105008}
  [\href{https://arxiv.org/abs/2112.12561}{{\ttfamily 2112.12561}}].

\bibitem{Arkani-Hamed:2020blm}
N.~Arkani-Hamed, T.-C.~Huang and Y.-t.~Huang, \emph{{The EFT-Hedron}},
  \href{https://doi.org/10.1007/JHEP05(2021)259}{\emph{JHEP} {\bfseries 05}
  (2021) 259} [\href{https://arxiv.org/abs/2012.15849}{{\ttfamily
  2012.15849}}].

\bibitem{Bellazzini:2016xrt}
B.~Bellazzini, \emph{{Softness and amplitudes\textquoteright{} positivity for
  spinning particles}},
  \href{https://doi.org/10.1007/JHEP02(2017)034}{\emph{JHEP} {\bfseries 02}
  (2017) 034} [\href{https://arxiv.org/abs/1605.06111}{{\ttfamily
  1605.06111}}].

\bibitem{deRham:2017avq}
C.~de~Rham, S.~Melville, A.J.~Tolley and S.-Y.~Zhou, \emph{{Positivity bounds
  for scalar field theories}},
  \href{https://doi.org/10.1103/PhysRevD.96.081702}{\emph{Phys. Rev. D}
  {\bfseries 96} (2017) 081702}
  [\href{https://arxiv.org/abs/1702.06134}{{\ttfamily 1702.06134}}].

\bibitem{deRham:2017imi}
C.~de~Rham, S.~Melville, A.J.~Tolley and S.-Y.~Zhou, \emph{{Massive Galileon
  Positivity Bounds}},
  \href{https://doi.org/10.1007/JHEP09(2017)072}{\emph{JHEP} {\bfseries 09}
  (2017) 072} [\href{https://arxiv.org/abs/1702.08577}{{\ttfamily
  1702.08577}}].

\bibitem{Camanho:2014apa}
X.O.~Camanho, J.D.~Edelstein, J.~Maldacena and A.~Zhiboedov, \emph{{Causality
  Constraints on Corrections to the Graviton Three-Point Coupling}},
  \href{https://doi.org/10.1007/JHEP02(2016)020}{\emph{JHEP} {\bfseries 02}
  (2016) 020} [\href{https://arxiv.org/abs/1407.5597}{{\ttfamily 1407.5597}}].

\bibitem{DAppollonio:2015fly}
G.~D'Appollonio, P.~Di~Vecchia, R.~Russo and G.~Veneziano, \emph{{Regge
  behavior saves String Theory from causality violations}},
  \href{https://doi.org/10.1007/JHEP05(2015)144}{\emph{JHEP} {\bfseries 05}
  (2015) 144} [\href{https://arxiv.org/abs/1502.01254}{{\ttfamily
  1502.01254}}].

\bibitem{deRham:2017zjm}
C.~de~Rham, S.~Melville, A.J.~Tolley and S.-Y.~Zhou, \emph{{UV complete me:
  Positivity Bounds for Particles with Spin}},
  \href{https://doi.org/10.1007/JHEP03(2018)011}{\emph{JHEP} {\bfseries 03}
  (2018) 011} [\href{https://arxiv.org/abs/1706.02712}{{\ttfamily
  1706.02712}}].

\bibitem{Kotanski:1970}
A.~Kotanski, \emph{{Transversity Amplitudes and Their Application to the Study
  of Collisions of Particle with Spin}}, {\emph{Acta Phys. Pol. B} {\bfseries
  1} (1970) 45}.

\bibitem{Wang:1966zza}
L.-L.C.~Wang, \emph{{General Method of Constructing Helicity Amplitudes Free
  from Kinematic Singularities and Zeros}},
  \href{https://doi.org/10.1103/PhysRev.142.1187}{\emph{Phys. Rev.} {\bfseries
  142} (1966) 1187}.

\bibitem{Cohen-Tannoudji:1968lnm}
G.~Cohen-Tannoudji, A.~Morel and H.~Navelet, \emph{{Kinematical singularities,
  crossing matrix and kinematical constraints for two-body helicity
  amplitudes}},
  \href{https://doi.org/10.1016/0003-4916(68)90243-1}{\emph{Annals Phys.}
  {\bfseries 46} (1968) 239}.

\bibitem{Veltman:1963th}
M.J.G.~Veltman, \emph{{Unitarity and causality in a renormalizable field theory
  with unstable particles}},
  \href{https://doi.org/10.1016/S0031-8914(63)80277-3}{\emph{Physica}
  {\bfseries 29} (1963) 186}.

\bibitem{Eden:1966dnq}
R.J.~Eden, P.V.~Landshoff, D.I.~Olive and J.C.~Polkinghorne, \emph{{The
  analytic S-matrix}}, Cambridge Univ. Press, Cambridge (1966).

\bibitem{Aoki:2022qbf}
K.~Aoki, \emph{{Unitarity and unstable-particle scattering amplitudes}},
  \href{https://doi.org/10.1103/PhysRevD.107.065017}{\emph{Phys. Rev. D}
  {\bfseries 107} (2023) 065017}
  [\href{https://arxiv.org/abs/2212.05670}{{\ttfamily 2212.05670}}].

\bibitem{Hannesdottir:2022bmo}
H.S.~Hannesdottir and S.~Mizera, \emph{{What is the i\ensuremath{\varepsilon}
  for the S-matrix?}}, SpringerBriefs in Physics, Springer (1, 2023),
  \href{https://doi.org/10.1007/978-3-031-18258-7}{10.1007/978-3-031-18258-7},
  [\href{https://arxiv.org/abs/2204.02988}{{\ttfamily 2204.02988}}].

\bibitem{Correia:2022dcu}
M.~Correia, \emph{{Nonperturbative Anomalous Thresholds}},
  \href{https://arxiv.org/abs/2212.06157}{{\ttfamily 2212.06157}}.

\bibitem{Nelson:2011sf}
A.E.~Nelson and J.~Scholtz, \emph{{Dark Light, Dark Matter and the Misalignment
  Mechanism}}, \href{https://doi.org/10.1103/PhysRevD.84.103501}{\emph{Phys.
  Rev. D} {\bfseries 84} (2011) 103501}
  [\href{https://arxiv.org/abs/1105.2812}{{\ttfamily 1105.2812}}].

\bibitem{Arias:2012az}
P.~Arias, D.~Cadamuro, M.~Goodsell, J.~Jaeckel, J.~Redondo and A.~Ringwald,
  \emph{{WISPy Cold Dark Matter}},
  \href{https://doi.org/10.1088/1475-7516/2012/06/013}{\emph{JCAP} {\bfseries
  06} (2012) 013} [\href{https://arxiv.org/abs/1201.5902}{{\ttfamily
  1201.5902}}].

\bibitem{Graham:2015rva}
P.W.~Graham, J.~Mardon and S.~Rajendran, \emph{{Vector Dark Matter from
  Inflationary Fluctuations}},
  \href{https://doi.org/10.1103/PhysRevD.93.103520}{\emph{Phys. Rev. D}
  {\bfseries 93} (2016) 103520}
  [\href{https://arxiv.org/abs/1504.02102}{{\ttfamily 1504.02102}}].

\bibitem{Christensen:2009jx}
N.D.~Christensen, P.~de~Aquino, C.~Degrande, C.~Duhr, B.~Fuks, M.~Herquet
  et~al., \emph{{A Comprehensive approach to new physics simulations}},
  \href{https://doi.org/10.1140/epjc/s10052-011-1541-5}{\emph{Eur. Phys. J. C}
  {\bfseries 71} (2011) 1541}
  [\href{https://arxiv.org/abs/0906.2474}{{\ttfamily 0906.2474}}].

\bibitem{Christensen:2008py}
N.D.~Christensen and C.~Duhr, \emph{{FeynRules - Feynman rules made easy}},
  \href{https://doi.org/10.1016/j.cpc.2009.02.018}{\emph{Comput. Phys. Commun.}
  {\bfseries 180} (2009) 1614}
  [\href{https://arxiv.org/abs/0806.4194}{{\ttfamily 0806.4194}}].

\bibitem{Alloul:2013bka}
A.~Alloul, N.D.~Christensen, C.~Degrande, C.~Duhr and B.~Fuks, \emph{{FeynRules
  2.0 - A complete toolbox for tree-level phenomenology}},
  \href{https://doi.org/10.1016/j.cpc.2014.04.012}{\emph{Comput. Phys. Commun.}
  {\bfseries 185} (2014) 2250}
  [\href{https://arxiv.org/abs/1310.1921}{{\ttfamily 1310.1921}}].

\bibitem{Hahn:2000kx}
T.~Hahn, \emph{{Generating Feynman diagrams and amplitudes with FeynArts 3}},
  \href{https://doi.org/10.1016/S0010-4655(01)00290-9}{\emph{Comput. Phys.
  Commun.} {\bfseries 140} (2001) 418}
  [\href{https://arxiv.org/abs/hep-ph/0012260}{{\ttfamily hep-ph/0012260}}].

\bibitem{Mertig:1990an}
R.~Mertig, M.~Bohm and A.~Denner, \emph{{FEYN CALC: Computer algebraic
  calculation of Feynman amplitudes}},
  \href{https://doi.org/10.1016/0010-4655(91)90130-D}{\emph{Comput. Phys.
  Commun.} {\bfseries 64} (1991) 345}.

\bibitem{Shtabovenko:2020gxv}
V.~Shtabovenko, R.~Mertig and F.~Orellana, \emph{{FeynCalc 9.3: New features
  and improvements}},
  \href{https://doi.org/10.1016/j.cpc.2020.107478}{\emph{Comput. Phys. Commun.}
  {\bfseries 256} (2020) 107478}
  [\href{https://arxiv.org/abs/2001.04407}{{\ttfamily 2001.04407}}].

\bibitem{Shtabovenko:2016sxi}
V.~Shtabovenko, R.~Mertig and F.~Orellana, \emph{{New Developments in FeynCalc
  9.0}}, \href{https://doi.org/10.1016/j.cpc.2016.06.008}{\emph{Comput. Phys.
  Commun.} {\bfseries 207} (2016) 432}
  [\href{https://arxiv.org/abs/1601.01167}{{\ttfamily 1601.01167}}].

\bibitem{Patel:2015tea}
H.H.~Patel, \emph{{Package-X: A Mathematica package for the analytic
  calculation of one-loop integrals}},
  \href{https://doi.org/10.1016/j.cpc.2015.08.017}{\emph{Comput. Phys. Commun.}
  {\bfseries 197} (2015) 276}
  [\href{https://arxiv.org/abs/1503.01469}{{\ttfamily 1503.01469}}].

\bibitem{Donnachie:1992ny}
A.~Donnachie and P.V.~Landshoff, \emph{{Total cross-sections}},
  \href{https://doi.org/10.1016/0370-2693(92)90832-O}{\emph{Phys. Lett. B}
  {\bfseries 296} (1992) 227}
  [\href{https://arxiv.org/abs/hep-ph/9209205}{{\ttfamily hep-ph/9209205}}].

\bibitem{Klusek:2009yi}
M.~Klusek, W.~Schafer and A.~Szczurek, \emph{{Exclusive production of rho0 rho0
  pairs in gamma gamma collisions at RHIC}},
  \href{https://doi.org/10.1016/j.physletb.2009.03.006}{\emph{Phys. Lett. B}
  {\bfseries 674} (2009) 92} [\href{https://arxiv.org/abs/0902.1689}{{\ttfamily
  0902.1689}}].

\bibitem{Reece:2018zvv}
M.~Reece, \emph{{Photon Masses in the Landscape and the Swampland}},
  \href{https://doi.org/10.1007/JHEP07(2019)181}{\emph{JHEP} {\bfseries 07}
  (2019) 181} [\href{https://arxiv.org/abs/1808.09966}{{\ttfamily
  1808.09966}}].

\bibitem{Hebecker:2017uix}
A.~Hebecker and P.~Soler, \emph{{The Weak Gravity Conjecture and the Axionic
  Black Hole Paradox}},
  \href{https://doi.org/10.1007/JHEP09(2017)036}{\emph{JHEP} {\bfseries 09}
  (2017) 036} [\href{https://arxiv.org/abs/1702.06130}{{\ttfamily
  1702.06130}}].

\bibitem{Arkani-Hamed:2006emk}
N.~Arkani-Hamed, L.~Motl, A.~Nicolis and C.~Vafa, \emph{{The String landscape,
  black holes and gravity as the weakest force}},
  \href{https://doi.org/10.1088/1126-6708/2007/06/060}{\emph{JHEP} {\bfseries
  06} (2007) 060} [\href{https://arxiv.org/abs/hep-th/0601001}{{\ttfamily
  hep-th/0601001}}].

\bibitem{Heidenreich:2016aqi}
B.~Heidenreich, M.~Reece and T.~Rudelius, \emph{{Evidence for a sublattice weak
  gravity conjecture}},
  \href{https://doi.org/10.1007/JHEP08(2017)025}{\emph{JHEP} {\bfseries 08}
  (2017) 025} [\href{https://arxiv.org/abs/1606.08437}{{\ttfamily
  1606.08437}}].

\bibitem{Heidenreich:2017sim}
B.~Heidenreich, M.~Reece and T.~Rudelius, \emph{{The Weak Gravity Conjecture
  and Emergence from an Ultraviolet Cutoff}},
  \href{https://doi.org/10.1140/epjc/s10052-018-5811-3}{\emph{Eur. Phys. J. C}
  {\bfseries 78} (2018) 337}
  [\href{https://arxiv.org/abs/1712.01868}{{\ttfamily 1712.01868}}].

\bibitem{Montero:2016tif}
M.~Montero, G.~Shiu and P.~Soler, \emph{{The Weak Gravity Conjecture in three
  dimensions}}, \href{https://doi.org/10.1007/JHEP10(2016)159}{\emph{JHEP}
  {\bfseries 10} (2016) 159}
  [\href{https://arxiv.org/abs/1606.08438}{{\ttfamily 1606.08438}}].

\bibitem{Arkani-Hamed:2005zuc}
N.~Arkani-Hamed, S.~Dimopoulos and S.~Kachru, \emph{{Predictive landscapes and
  new physics at a TeV}},
  \href{https://arxiv.org/abs/hep-th/0501082}{{\ttfamily hep-th/0501082}}.

\bibitem{Dvali:2007hz}
G.~Dvali, \emph{{Black Holes and Large N Species Solution to the Hierarchy
  Problem}}, \href{https://doi.org/10.1002/prop.201000009}{\emph{Fortsch.
  Phys.} {\bfseries 58} (2010) 528}
  [\href{https://arxiv.org/abs/0706.2050}{{\ttfamily 0706.2050}}].

\bibitem{Dvali:2007wp}
G.~Dvali and M.~Redi, \emph{{Black Hole Bound on the Number of Species and
  Quantum Gravity at LHC}},
  \href{https://doi.org/10.1103/PhysRevD.77.045027}{\emph{Phys. Rev. D}
  {\bfseries 77} (2008) 045027}
  [\href{https://arxiv.org/abs/0710.4344}{{\ttfamily 0710.4344}}].

\bibitem{Hoskins:1985tn}
J.K.~Hoskins, R.D.~Newman, R.~Spero and J.~Schultz, \emph{{Experimental tests
  of the gravitational inverse square law for mass separations from 2-cm to
  105-cm}}, \href{https://doi.org/10.1103/PhysRevD.32.3084}{\emph{Phys. Rev. D}
  {\bfseries 32} (1985) 3084}.

\bibitem{Smith:1999cr}
G.L.~Smith, C.D.~Hoyle, J.H.~Gundlach, E.G.~Adelberger, B.R.~Heckel and
  H.E.~Swanson, \emph{{Short range tests of the equivalence principle}},
  \href{https://doi.org/10.1103/PhysRevD.61.022001}{\emph{Phys. Rev. D}
  {\bfseries 61} (2000) 022001}.

\bibitem{Kapner:2006si}
D.J.~Kapner, T.S.~Cook, E.G.~Adelberger, J.H.~Gundlach, B.R.~Heckel, C.D.~Hoyle
  et~al., \emph{{Tests of the gravitational inverse-square law below the
  dark-energy length scale}},
  \href{https://doi.org/10.1103/PhysRevLett.98.021101}{\emph{Phys. Rev. Lett.}
  {\bfseries 98} (2007) 021101}
  [\href{https://arxiv.org/abs/hep-ph/0611184}{{\ttfamily hep-ph/0611184}}].

\bibitem{Wagner:2012ui}
T.A.~Wagner, S.~Schlamminger, J.H.~Gundlach and E.G.~Adelberger,
  \emph{{Torsion-balance tests of the weak equivalence principle}},
  \href{https://doi.org/10.1088/0264-9381/29/18/184002}{\emph{Class. Quant.
  Grav.} {\bfseries 29} (2012) 184002}
  [\href{https://arxiv.org/abs/1207.2442}{{\ttfamily 1207.2442}}].

\bibitem{Lee:2020zjt}
J.G.~Lee, E.G.~Adelberger, T.S.~Cook, S.M.~Fleischer and B.R.~Heckel,
  \emph{{New Test of the Gravitational $1/r^2$ Law at Separations down to 52
  $\mu$m}}, \href{https://doi.org/10.1103/PhysRevLett.124.101101}{\emph{Phys.
  Rev. Lett.} {\bfseries 124} (2020) 101101}
  [\href{https://arxiv.org/abs/2002.11761}{{\ttfamily 2002.11761}}].

\bibitem{Yang:2012zzb}
S.-Q.~Yang, B.-F.~Zhan, Q.-L.~Wang, C.-G.~Shao, L.-C.~Tu, W.-H.~Tan et~al.,
  \emph{{Test of the Gravitational Inverse Square Law at Millimeter Ranges}},
  \href{https://doi.org/10.1103/PhysRevLett.108.081101}{\emph{Phys. Rev. Lett.}
  {\bfseries 108} (2012) 081101}.

\bibitem{Tan:2020vpf}
W.-H.~Tan et~al., \emph{{Improvement for Testing the Gravitational
  Inverse-Square Law at the Submillimeter Range}},
  \href{https://doi.org/10.1103/PhysRevLett.124.051301}{\emph{Phys. Rev. Lett.}
  {\bfseries 124} (2020) 051301}.

\bibitem{Chen:2014oda}
Y.J.~Chen, W.K.~Tham, D.E.~Krause, D.~Lopez, E.~Fischbach and R.S.~Decca,
  \emph{{Stronger Limits on Hypothetical Yukawa Interactions in the
  30\textendash{}8000 nm Range}},
  \href{https://doi.org/10.1103/PhysRevLett.116.221102}{\emph{Phys. Rev. Lett.}
  {\bfseries 116} (2016) 221102}
  [\href{https://arxiv.org/abs/1410.7267}{{\ttfamily 1410.7267}}].

\bibitem{MICROSCOPE:2022doy}
{\scshape MICROSCOPE} collaboration, \emph{{MICROSCOPE Mission: Final Results
  of the Test of the Equivalence Principle}},
  \href{https://doi.org/10.1103/PhysRevLett.129.121102}{\emph{Phys. Rev. Lett.}
  {\bfseries 129} (2022) 121102}
  [\href{https://arxiv.org/abs/2209.15487}{{\ttfamily 2209.15487}}].

\bibitem{Hardy:2016kme}
E.~Hardy and R.~Lasenby, \emph{{Stellar cooling bounds on new light particles:
  plasma mixing effects}},
  \href{https://doi.org/10.1007/JHEP02(2017)033}{\emph{JHEP} {\bfseries 02}
  (2017) 033} [\href{https://arxiv.org/abs/1611.05852}{{\ttfamily
  1611.05852}}].

\bibitem{Escudero:2019gvw}
M.~Escudero and S.J.~Witte, \emph{{A CMB search for the neutrino mass mechanism
  and its relation to the Hubble tension}},
  \href{https://doi.org/10.1140/epjc/s10052-020-7854-5}{\emph{Eur. Phys. J. C}
  {\bfseries 80} (2020) 294}
  [\href{https://arxiv.org/abs/1909.04044}{{\ttfamily 1909.04044}}].

\bibitem{Ibe:2020dly}
M.~Ibe, S.~Kobayashi, Y.~Nakayama and S.~Shirai, \emph{{Cosmological Constraint
  on Vector Mediator of Neutrino-Electron Interaction in light of XENON1T
  Excess}}, \href{https://doi.org/10.1007/JHEP12(2020)004}{\emph{JHEP}
  {\bfseries 12} (2020) 004}
  [\href{https://arxiv.org/abs/2007.16105}{{\ttfamily 2007.16105}}].

\bibitem{XENON:2022ltv}
{\scshape XENON} collaboration, \emph{{Search for New Physics in Electronic
  Recoil Data from XENONnT}},
  \href{https://doi.org/10.1103/PhysRevLett.129.161805}{\emph{Phys. Rev. Lett.}
  {\bfseries 129} (2022) 161805}
  [\href{https://arxiv.org/abs/2207.11330}{{\ttfamily 2207.11330}}].

\bibitem{A:2022acy}
{{ShivaSankar K.A.}}, A.~Majumdar, D.K.~Papoulias, H.~Prajapati and
  R.~Srivastava, \emph{{Implications of first LZ and XENONnT results: A
  comparative study of neutrino properties and light mediators}},
  \href{https://doi.org/10.1016/j.physletb.2023.137742}{\emph{Phys. Lett. B}
  {\bfseries 839} (2023) 137742}
  [\href{https://arxiv.org/abs/2208.06415}{{\ttfamily 2208.06415}}].

\bibitem{Ellis:2016jkw}
J.~Ellis, \emph{{TikZ-Feynman: Feynman diagrams with TikZ}},
  \href{https://doi.org/10.1016/j.cpc.2016.08.019}{\emph{Comput. Phys. Commun.}
  {\bfseries 210} (2017) 103}
  [\href{https://arxiv.org/abs/1601.05437}{{\ttfamily 1601.05437}}].

\bibitem{Dohse:2018vqo}
M.~Dohse, \emph{{TikZ-FeynHand: Basic User Guide}},
  \href{https://arxiv.org/abs/1802.00689}{{\ttfamily 1802.00689}}.

\end{thebibliography}\endgroup

\end{document}